\documentclass[11pt,a4paper]{article}
\pdfoutput=1
\usepackage{jheppub}
\usepackage{amsmath,amssymb,amsfonts}
\usepackage{graphicx,graphics}
\usepackage{graphics}
\usepackage{url}
\usepackage{slashed}
\usepackage{color}
\usepackage{mathtools}
\usepackage[utf8]{inputenc}
\usepackage{microtype} %% Fancier typesetting

\title{Revisiting the D-meson hadroproduction in general-mass variable flavour number scheme}

\affiliation[a]{Institute for Theoretical Physics, T\"ubingen University, Auf der Morgenstelle 14, 72076 T\"ubingen, Germany}
\affiliation[b]{University of Jyvaskyla, Department of Physics, P.O. Box 35, FI-40014 University of Jyvaskyla, Finland}
\affiliation[c]{Helsinki Institute of Physics, P.O. Box 64, FI-00014 University of Helsinki, Finland}

\emailAdd{hannu.paukkunen@jyu.fi}
\emailAdd{ilkka.helenius@uni-tuebingen.de}

\abstract{
We introduce a novel realization of the open heavy-flavour hadroproduction in general-mass variable flavour number scheme at next-to-leading order in perturbative QCD. The principal novelty with respect to the earlier works is in the treatment of small-transverse-momentum limit, which has been a particularly challenging kinematic region in the past. We show that by a suitable choice of scheme, it is possible to obtain a well-behaved description of the open heavy-flavour hadroproduction cross sections from zero up to asymptotically high transverse momentum. We contrast our calculation with the available D$^0$-meson data as measured by the LHCb and ALICE collaborations at the LHC, finding a very good agreement within the theoretical and experimental uncertainties. We also compare our framework with other theoretical approaches.
}


\begin{document}

\author{Ilkka Helenius$^a$ and}
\author{Hannu Paukkunen$^{b,c}$}

\maketitle

\section{Introduction}
\label{sec:Introduction}

The hadroproduction of heavy-flavoured mesons at the LHC, in particular the D- and B-meson measurements at forward direction \cite{Aaij:2013mga,Aaij:2015bpa,Aaij:2016jht,Aaij:2017qml}, has recently attracted a growing interest for its potential to provide information on partonic dynamics at low momentum fractions. Because of the finite heavy-quark mass $m$, the perturbative methods are applicable down to zero transverse momentum ($p_{\rm T}$) of the observed meson, and the measurements provide opportunities e.g. to constrain the collinearly factorized gluon distributions at small momentum fractions in proton \cite{Cacciari:2015fta, Zenaiev:2015rfa, Gauld:2016kpd} or nucleus \cite{Gauld:2015lxa, Kusina:2017gkz}, or to test other scenarios like saturation physics \cite{Ducloue:2016ywt,Fujii:2017rqa}, or $k_{\rm T}$ factorization \cite{Maciula:2013wg}. The D-meson production is also of great interest from the viewpoint of neutrino astrophysics as the secondary neutrinos from D mesons produced in scatterings of cosmic rays in the atmosphere form a significant background for the extraterrestrial neutrinos. Given that the D-meson measurements at LHCb \cite{Aaij:2013mga,Aaij:2015bpa,Aaij:2016jht} are kinematically close to the cosmic-ray-on-air scattering, the rates for secondary neutrinos can be constrained by the LHC data \cite{Gauld:2015kvh,Bhattacharya:2016jce,Benzke:2017yjn,Garzelli:2016xmx}. In heavy-ion collisions the measured open heavy-flavour data \cite{Adam:2015sza,Sirunyan:2017xss} provides opportunities e.g. to test the so-called dead-cone effect \cite{Dokshitzer:1991fd,Dokshitzer:2001zm} in QCD medium \cite{Prino:2016cni,Andronic:2015wma}.

Theoretically, there are several collinear-factorization-based ways to calculate cross sections for heavy-flavoured mesons in proton-proton (p-p) collisions, see e.g. Refs.~\cite{Klasen:2014dba,Zenaiev:2016kfl,Andronic:2015wma} for reviews. On one hand, parton-level heavy-quark cross sections at fixed-flavour-number scheme (FFNS) \cite{Nason:1989zy,Beenakker:1990maa,Bojak:2001fx} can be folded with phenomenological, scale-independent parton-to-meson fragmentation functions (FFs), or the parton-level calculation is matched to a parton-shower \cite{Frixione:2002ik, Frixione:2007vw} from a general-purpose Monte-Carlo event generator, such as \textsc{Pythia 8} \cite{Sjostrand:2014zea} or \textsc{Herwig} \cite{Bahr:2008pv}, and the showered event is then hadronized according to the hadronization model of the generator. Alternatively, one can work fully within the collinear factorization where the fragmentation is described with universal, scale-dependent FFs \cite{Albino:2008gy}. In this paper, we will focus on this latter approach.
 
The general framework in QCD to treat the heavy-quark production is the so-called general-mass variable flavour number scheme (GM-VFNS) \cite{Collins:1998rz,Thorne:2008xf,Forte:2010ta,Tung:2001mv,Aivazis:1993pi,Olness:1997yc,Cacciari:1998it}. In this framework, at low interaction scales $Q^2 \lesssim m^2$ the heavy quarks are not treated as partons in PDFs but are considered only as massive objects in the final state. The full mass dependence is retained in the production cross sections, but the initial-state partons are restricted only to the light ones. These cross sections contain mass-dependent logarithmic terms which, towards higher interaction scales, will eventually dominate and diverge. In GM-VFNS these large logarithms are subtracted at a certain transition scale $Q_{\rm t}^2$ -- typically the heavy-quark mass threshold -- and resummed into the PDFs and scale-dependent FFs. At asymptotically high interaction scales $Q^2 \gg m^2$ the result reduces (up to finite terms) to the calculation where the quark mass has been put to zero from the outset, the so-called zero-mass variable flavour number scheme (ZM-VFNS).

To obtain a well-behaved description for the heavy-flavoured mesons within GM-VFNS approach from zero to asymptotically large $p_{\rm T}$ has, however, been a bit challenging. The difficulty is related to the intrinsic freedom in GM-VFNS to use the zero-mass formalism for the processes with heavy-quarks in the initial state or where the fragmenting parton is a light one. The massless coefficient functions display a divergent behaviour towards low $p_{\rm T}$ and with a typical scale choice $Q^2 \sim {m^2 + p_{\rm T}^2}$ their contribution dominates the cross sections immediately above $p_{\rm T}=0$. Thus, the production cross sections diverge towards $p_{\rm T} \rightarrow 0$. 

A solution was proposed in Ref.~\cite{Kniehl:2015fla}. In essence, the idea was to exclude the aforementioned divergent contributions at low $p_{\rm T}$ by retaining the factorization and fragmentation scales at the threshold $Q^2=m^2$ until large-enough $p_{\rm T}$. Formally, the difference with respect to a more natural choice $Q^2 = m^2 + p_{\rm T}^2$ is one order higher in QCD coupling than what one works at, but numerically the effect is large and the cross sections are rendered finite down to $p_{\rm T}=0$. A relatively good description of the LHCb data can be obtained by tuning the scales in this manner \cite{Kniehl:2015fla,Benzke:2017yjn}, but the price to pay is that there will be a certain unphysical wiggle in the production cross section near the region where one decides to turn on the heavy-quark PDFs and light-parton FFs, see e.g. Figure~6 in Ref.~\cite{Kniehl:2015fla}. An alternative strategy along this line would be to take the transition scale $Q_{\rm t}^2$ to be much higher than the heavy-quark mass \cite{Bertone:2017djs}. However, this would lead to a discontinuity in the cross sections at the arbitrary point where one decides to make the transition. Higher-order calculations should decrease the transient effects in both cases, but will not cure them completely. Clearly, a different solution would be beneficial.

The option we propose here is to make use of the scheme dependence inherent to GM-VFNS. Physically, our choice of scheme is rooted in the observation that --- in the absence of intrinsic charm component --- the contributions from heavy-quark PDFs and light-parton FFs are simply an efficient way to resum diagrams where a heavy quark-antiquark ($Q\overline{Q}$) pair is dynamically produced. Being of the same origin, it is natural to require that these contributions respect the same kinematical constraints as the channels where the pair is explicit produced. These are formally $\mathcal{O}(m^2)$ effects and can be included in the definition of a scheme. However, the contributions from heavy-quark PDFs and light-parton FFs will no longer diverge in the $p_{\rm T} \rightarrow 0$ limit, but are regulated by the heavy-quark mass. The production cross sections thus remain finite in the $p_{\rm T} \rightarrow 0$ limit with arbitrary factorization and fragmentation scales.

\section{Formalism}
\label{sec:Formalism}

In this section we will describe our theoretical construction and its numerical implementation. As the GM-VFNS framework in hadroproduction of heavy quarks has been detailedly discussed in Refs.~\cite{Olness:1997yc,Kniehl:2004fy,Kniehl:2005mk}, we will here focus only on the most important features of our approach. However, enough details are still given so that our results can be reproduced.

\subsection{General structure and kinematics}
\label{sec:Generalstructureandkinematics}

The process we study is an inclusive production of a hadron $h_3$ with momentum $P_3$ in collision of two hadrons $h_1$ and $h_2$ with momenta $P_1$ and $P_2$,
$$
h_1(P_1) + h_2(P_2) \rightarrow h_3(P_3) + X \, .
$$
In the approximation where the masses of partons and produced hadron are neglected, the cross section differentiated with respect to the produced hadron's transverse momentum $P_{\rm T}$ and rapidity $Y$ can be written in the well-known factorized form \cite{Aversa:1988vb},
\begin{align}
& \frac{\mathrm{d}\sigma(h_1 + h_2 \rightarrow h_3 + X)}{\mathrm{d}P_{\rm T}\mathrm{d}Y} =
 \sum _{ijk} \int_{z^{\rm min}}^1 \frac{\mathrm{d}z}{z} \int_{x_1^{\rm min}}^1 \mathrm{d}x_1 \int _{x_2^{\rm min}}^1 \mathrm{d}x_2 \nonumber \\
& D_{k \rightarrow h_3}(z,\mu^2_{\rm frag}) \, 
f_i^{h_1}(x_1,\mu^2_{\rm fact}) f_j^{h_2}(x_2,\mu^2_{\rm fact}) \frac{\mathrm{d}\hat{\sigma}^{ij\rightarrow k}(\tau_1, \tau_2, \mu^2_{\rm ren}, \mu^2_{\rm fact}, \mu^2_{\rm frag})}{\mathrm{d}p_{\rm T} \mathrm{d}y}
\label{eq:masterZM} .
\end{align}
where the fragmenting parton's transverse momentum and rapidity are $p_{\rm T} = P_{\rm T}/z$ and $y=Y$. Here $f_i^{h_{1,2}}(x_{1,2},\mu^2_{\rm fact})$ are the PDFs for parton species $i$ in hadron $h_{1,2}$ and $D_{l \rightarrow h_3}(z,\mu^2_{\rm frag})$ is the parton-to-$h_{3}$ FF. The invariants $\tau_i$ are defined as
\begin{equation}
 \tau_1 \equiv \frac{p_1 \cdot p_3}{p_1 \cdot p_2} = \frac{p_{\rm T} \, e^{-y}}{x_2 \sqrt{s}}, \quad
 \tau_2 \equiv \frac{p_2 \cdot p_3}{p_1 \cdot p_2} = \frac{p_{\rm T} \, e^{y}}{x_1 \sqrt{s}} \,,
\end{equation}
where $p_1$ and $p_2$ are the momenta of the incoming partons, $p_3$ is the momentum of the produced, outgoing parton and $\sqrt{s}$ is center-of-mass (c.m.) energy of the collision. The integration limits are given by

\begin{equation}
  x_1^{\rm min} = \frac{p_{\rm T} \, e^{ y}}{\sqrt{s}-p_{\rm T} \, e^{-y}}, \quad
  x_2^{\rm min} = \frac{x_1 p_{\rm T} \, e^{-y}}{x_1\sqrt{s}-p_{\rm T} \, e^{y}}, \quad 
  z^{\rm min}   = \frac{2P_{\rm T}\cosh Y}{\sqrt{s}} \, . \label{eq:ZMmin}
 \end{equation}

\subsubsection{Partonic kinematics in the presence of mass}
\label{sec:Partonickinematicsinthepresenceofmass} 
 
When a $Q\overline{Q}$ pair is produced from light partons, the zero-mass partonic kinematics above should be adjusted to account for the heavy-quark mass $m$. In practice, this amounts to replacing the partonic transverse momentum $p_{\rm T}$ in the $x_{1,2}$ integration limits and scaling variables $\tau_{1,2}$ by the partonic transverse mass $m_{\rm T} \equiv \sqrt{p_{\rm T}^2+m^2}$ ,
\begin{equation}
  x_1^{\rm min} \rightarrow \frac{m_{\rm T} \, e^{ y}}{\sqrt{s}-m_{\rm T} \, e^{-y}}, \quad
  x_2^{\rm min} \rightarrow \frac{x_1 m_{\rm T} \, e^{-y}}{x_1\sqrt{s}-m_{\rm T} \, e^{y}}, \quad 
  \tau_1 \rightarrow \frac{m_{\rm T} \, e^{-y}}{x_2 \sqrt{s}} , \quad 
  \tau_2 \rightarrow \frac{m_{\rm T} \, e^{ y}}{x_1 \sqrt{s}}. \label{eq:massivekin}
\end{equation}
These kinematics correspond to the inclusive heavy-quark production. When the produced parton is a heavy quark, the above replacements follow directly from the momentum conservation. However, in the case that the fragmenting parton is a light one or when there is a heavy quark in the initial state, these replacements are strictly speaking not necessary, but are part of our choice of scheme (SACOT-$m_{\rm T}$, explained in more detail later). In the picture where the heavy quarks are generated perturbatively, the heavy-flavour PDFs and light-flavour FFs are merely an efficient way to resum diagrams where a heavy quark-antiquark pair is created. That is, the production of heavy-flavour pair is implicit in these contributions and motivates the usage of heavy-flavour kinematics.

\subsubsection{Massive fragmentation variable}
\label{sec:Massivefragmentationvariable} 

The zero-mass version of the fragmentation scaling variable $z=P_3/p_3$ is ill-defined in the presence of massive quarks/hadrons, and the zero-mass relations $Y=y$ and $P_{\rm T} = z p_{\rm T}$ are no longer true. Here, we choose to define the scaling variable $z$ in a Lorentz-invariant way as
\begin{equation}
 z \equiv \frac{P_3 \cdot \left(P_1 + P_2 \right)}{p_3 \cdot \left(P_1 + P_2 \right)} \xrightarrow{\rm hadronic \,\, c.m. \,\, frame} \frac{E_{\rm hadron}}{E_{\rm parton}} \, . \label{eq:massivez}
\end{equation}
As indicated, in the c.m. frame of the colliding hadrons $z$ can be interpreted as the fraction of partonic energy carried by the outgoing hadron \cite{Eskola:2002kv}. Alternatively, the scaling variable could be defined e.g. in terms of light-cone momentum fractions \cite{Albino:2008fy,Kniehl:2015fla}. From the above definition and considering the fragmentation to be collinear in the c.m. frame, we have two equations,
\begin{equation}
 z = \frac{M_{\rm T} \cosh Y}{m_{\rm T} \cosh y}, \quad \frac{P_{\rm T}}{M_{\rm T} \sinh Y} = \frac{p_{\rm T}}{m_{\rm T} \sinh y},
\end{equation}
where the hadronic transverse mass is defined as $M_{\rm T} \equiv \sqrt{M^2+P_{\rm T}^2}$, $M$ being the mass of the produced hadron. We can solve these equations for the hadronic transverse momentum and rapidity,
\begin{align}
P_{\rm T}^2(y,p_{\rm T}) & = \frac{z^2 m^2_{\rm T}\cosh^2 y - M^2}{1 + ({m_{\rm T}^2\sinh^2 y})/{p^2_{\rm T}}},  \\
    Y(y,p_{\rm T}) & = \sinh^{-1} \left(\frac{m_{\rm T} \sinh y}{p_{\rm T}} \frac{P_{\rm T}}{M_{\rm T}}\right) \, .
\end{align}
The cross section corresponding to the above definition of $z$ can be obtained as
\begin{align}
& \frac{\mathrm{d}\sigma(h_1 + h_2 \rightarrow h_3 + X)}{\mathrm{d}P_{\rm T}dY} =
 \sum _{ijk} \int_{z^{\rm min}}^1 \mathrm{d}z \int_{x_1^{\rm min}}^1 \mathrm{d}x_1 \int_{x_2^{\rm min}}^1 \mathrm{d}x_2 \int \mathrm{d}y \int \mathrm{d}p_{\rm T} \nonumber \\
& D_{k \rightarrow h_3}(z,\mu^2_{\rm frag}) \, 
f_i^{h_1}(x_1,\mu^2_{\rm fact}) f_j^{h_2}(x_2,\mu^2_{\rm fact}) \frac{\mathrm{d}\hat{\sigma}^{ij\rightarrow k}(\tau_1, \tau_2, m, \mu^2_{\rm ren}, \mu^2_{\rm fact}, \mu^2_{\rm frag})}{\mathrm{d}p_{\rm T} \mathrm{d}y} \\
& \delta \left(Y-Y(y,p_{\rm T})\right) \delta \left(P_{\rm T}-P_{\rm T}(y,p_{\rm T})\right) \nonumber
\end{align}
by integrating over $p_{\rm T}$ and $y$. Using the relation
\begin{equation}
 \int \mathrm{d}y\mathrm{d}p_{\rm T} = \frac{1}{z} \int \mathrm{d}P_{\rm T}(y,p_{\rm T})\mathrm{d}Y(y,p_{\rm T}) \,,
\end{equation}
we find again Eq.~(\ref{eq:masterZM})
where the partonic transverse momentum and rapidity are now given by
\begin{align}
p_{\rm T}^2 & = \frac{M^2_{\rm T}\cosh^2 Y - z^2m^2}{z^2} \left(1 + \frac{M_{\rm T}^2\sinh^2 Y}{P^2_{\rm T}}\right)^{-1} \, , \label{eq:pt_PT} \\
    y & = \sinh^{-1} \left(\frac{M_{\rm T} \sinh Y}{P_{\rm T}} \frac{p_{\rm T}}{m_{\rm T}}\right) \, , \label{eq:y_Y}
\end{align}
and the hadron mass corrects the lower limit of the $z$ integration as
\begin{equation}
z^{\rm min}   = \frac{2M_{\rm T}\cosh Y}{\sqrt{s}} \, . 
\end{equation}
Otherwise the cross-section formula is formally identical to the case of zero-mass partons and hadrons.

\subsection{Partonic cross sections in SACOT and SACOT-$m_{\rm T}$ schemes}
\label{sec:Partoniccrosssections}

The starting point in our GM-VFNS construction, is the next-to-leading-order (NLO) one-particle inclusive heavy-quark cross section in FFNS \cite{Nason:1989zy,Beenakker:1990maa,Bojak:2001fx} where heavy flavour can be produced in three different partonic processes,
\begin{equation}
g+g \rightarrow Q + X, \quad q+\overline{q} \rightarrow Q + X, \quad q+g \rightarrow Q + X \,. \label{eq:QQbarproc}
\end{equation}
In FFNS (and also GM-VFNS at low interaction scales), these are the only ways to produce heavy flavour. The heavy-quark mass $m$ is kept finite in these processes and in the high-$p_{\rm T}$ limit, the partonic cross sections develop logarithmic divergences $\sim\log(p_{\rm T}^2/m^2)$ coming from kinematic regions where the heavy quarks become collinear with other partons. These are the first terms in the whole series of large collinear logarithms which, in GM-VFNS framework, are resummed to heavy-quark PDFs and parton-to-hadron FFs when the interaction scale exceeds a chosen transition scale $Q_{\rm t}$. From now on identify $Q_{\rm t}$ as the heavy quark mass, $Q_{\rm t} =m$. To avoid double counting when including the contributions also from heavy-quark PDFs and using the scale-dependent parton-to-hadron FFs, one has then to subtract these logarithmic pieces from the coefficient functions. In what follows, we will explain what are the added and subtracted terms in our case, using the $g+g \rightarrow (Q \rightarrow h_3) + X$ channel as an explicit example.

\begin{figure}[tbhp]
\centering
\includegraphics[width=1.00\textwidth]{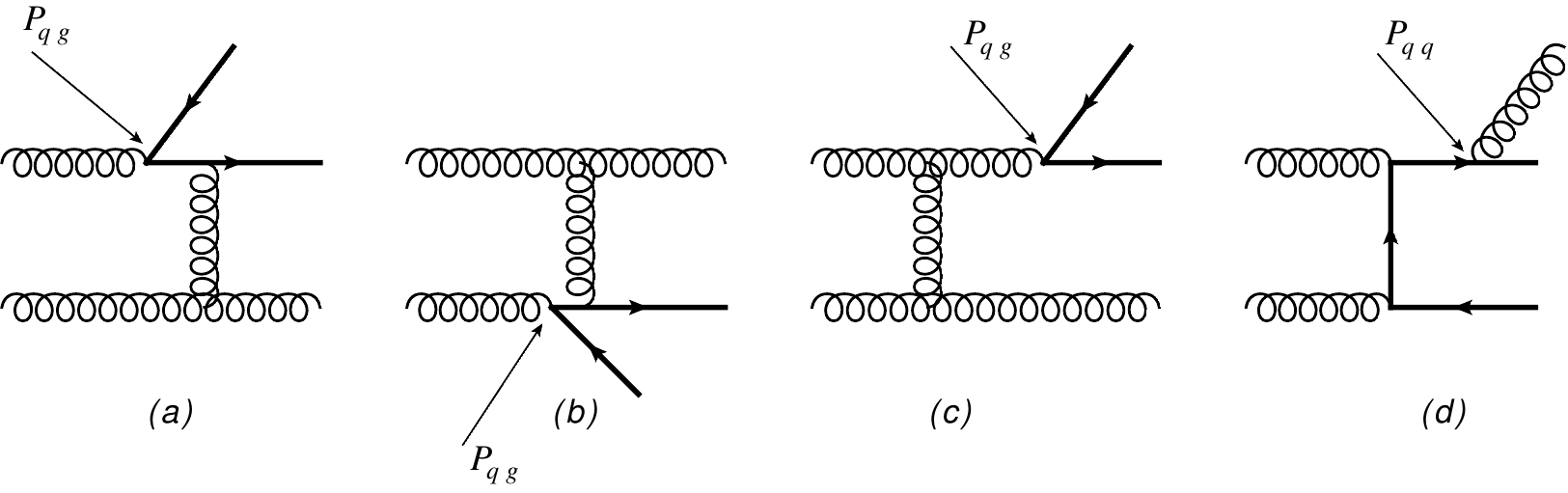}
\caption{Diagrams illustrating the origin of four types of collinear logarithms in $g+g \rightarrow Q + X \rightarrow h_3 + X$ channel. The relevant partonic splitting functions are indicated. Made with JaxoDraw 2.1 \cite{Binosi:2003yf}.} 
\label{fig:fig1}
\end{figure}

The gluon-fusion process $g+g \rightarrow (Q \rightarrow h_3) + X$ at NLO entails four different sources of collinear divergences in the $m \rightarrow 0$ (or equivalently $p_{\rm T} \rightarrow \infty$) limit, illustrated in Figure~\ref{fig:fig1}: \\ 
--- one of the two initial-state gluons splits into a collinear heavy quark-antiquark pair, \\
--- an outgoing gluon splits into collinear heavy quark-antiquark pair, \\
--- an outgoing heavy quark emits a collinear gluon. \\
A simple way to specify the GM-VFNS subtraction terms at NLO is to take as the starting point the leading-order (LO) contributions from channels where there are heavy quarks in the initial state or the fragmenting parton is a light one. Let us begin with the former case. Using Eq.~(\ref{eq:masterZM}), we write the leading-order contribution for process $g+Q\rightarrow (Q \rightarrow h_3) + X$ as
\begin{align}
\mathrm{d}\sigma^{gQ\rightarrow (Q \rightarrow h_3)+X}_{\rm LO} = & \int_{z^{\rm min}}^1 \frac{\mathrm{d}z}{z} \int_{x_1^{\rm min}}^1 \mathrm{d}x_1  \, \frac{p_{\rm T}}{8\pi s^2} \frac{f_g^{h_1}(x_1,\mu^2_{\rm fact})}{x_1} \frac{f_Q^{h_2}(x_2^{\rm min},\mu^2_{\rm fact})}{x_2^{\rm min}} \label{eq:gQ-Q} \\
& D_{Q \rightarrow h_3}(z,\mu^2_{\rm frag}) \frac{|\mathcal{M}_{gQ\rightarrow Qg}|^2}{x_1(1-\tau_2)} \,. \nonumber
\end{align}
This now uniquely determines the subtraction term which cancels the logarithmic term from diagrams like (b) in Figure~\ref{fig:fig1}. The expression for perturbative heavy-quark PDF, to the first order in strong coupling $\alpha_s$, reads
\begin{equation}
f_Q(x,\mu^2_{\rm fact}) = \left(\frac{\alpha_s}{2\pi}\right) \log\left(\frac{\mu^2_{\rm fact}}{m^2}\right) \int_x^1 \frac{\mathrm{d}\ell}{\ell} P_{qg}\left(\frac{x}{\ell}\right) f_g(\ell,\mu^2_{\rm fact}) + \mathcal{O}(\alpha_s^2) \,, \label{eq:fQ}
\end{equation}
where $P_{qg}(z)=T_f\left[z^2+(1-z)^2\right]$ with $T_f=1/2$, is the leading-order gluon-to-quark splitting function. Using this expression for $f_Q^{h_2}$ in Eq.~(\ref{eq:gQ-Q}) gives our definition of the subtraction term,
\begin{align}
S^{gQ\rightarrow (Q \rightarrow h_3)+X} = & \left(\frac{\alpha_s}{2\pi}\right) \log\left(\frac{\mu^2_{\rm fact}}{m^2}\right) \int_{z^{\rm min}}^1 \frac{\mathrm{d}z}{z} \int_{x_1^{\rm min}}^1 \mathrm{d}x_1  \, \frac{p_{\rm T}}{8\pi s^2} \frac{f_g^{h_1}(x_1,\mu^2_{\rm fact})}{x_1} \label{eq:subt1} \\ 
& \frac{1}{x_2^{\rm min}} \int_{x_2^{\rm min}}^1 \frac{\mathrm{d}\ell}{\ell} P_{qg}\left(\frac{x_2^{\rm min}}{\ell}\right) f_g^{h_2}(\ell,\mu^2_{\rm fact}) \nonumber  D_{Q \rightarrow h_3}(z,\mu^2_{\rm frag}) \frac{|\mathcal{M}_{gQ\rightarrow Qg}|^2}{x_1(1-\tau_2)} \,. 
\end{align}
When adding the leading-order contribution of Eq.~(\ref{eq:gQ-Q}), one must then compensate by subtracting Eq.~(\ref{eq:subt1}). The difference contributes at $\mathcal{O}(\alpha_s^4)$ and is not considered at an NLO-level $\mathcal{O}(\alpha_s^3)$ calculation. Here, we also plainly see the origin of the scheme dependence in GM-VFNS: The exact form of $|\mathcal{M}_{gQ\rightarrow Qg}|^2$ appearing in Eq.~(\ref{eq:gQ-Q}) and Eq.~(\ref{eq:subt1}) is subject to a certain amount of arbitrariness. Indeed, the only requirement is that in the $m \rightarrow 0$ limit $|\mathcal{M}_{gQ\rightarrow Qg}|^2$ must tend to the zero-mass expression $|\mathcal{M}_{gq\rightarrow qg}|^2$ so as to ensure that the corresponding collinear logarithm from $g+g \rightarrow (Q \rightarrow h_3) + X$ process cancels.
Otherwise we can choose it at will. Similarly, the exact expressions for the integration limits are irrelevant as far as the zero-mass expressions given in Eq.~(\ref{eq:ZMmin}) are found in the $m \rightarrow 0$ limit. The simplest option is to use the zero-mass matrix elements and kinematics from the beginning --- this choice of scheme is usually dubbed as simplified ACOT, or SACOT scheme \cite{Kramer:2000hn}. Here, we shall adopt a prescription where we use the zero-mass matrix elements but still retain the kinematic mass dependence. In other words, the integration limits and the invariants $\tau_1$ and $\tau_2$ are as in Eq.~(\ref{eq:massivekin}), and for the squared matrix element in Eqs.~(\ref{eq:gQ-Q}) and (\ref{eq:subt1}) we take,
\begin{equation}
|\mathcal{M}_{gQ\rightarrow Qg}|^2 = |\mathcal{M}_{gq\rightarrow qg}|^2(\tau_1,\tau_2) = 16\pi^2\alpha_s^2 \frac{2C_f}{D_A} \left(1 + \tau_1^2 \right) \left(\frac{C_A}{\tau_2^2} + \frac{C_f}{\tau_1} \right) ,
\end{equation}
where $|\mathcal{M}_{gq\rightarrow qg}|^2(\tau_1,\tau_2)$ is obtained from the zero-mass expression \cite{Ellis:1985er},
\begin{equation}
|\mathcal{M}_{gq\rightarrow qg}|^2 = 16\pi^2\alpha_s^2 \left(\hat s^2 + \hat t^2 \right) \left(\frac{C_A}{\hat u^2} + \frac{C_f}{\hat s \hat t} \right) \frac{2C_f}{D_A},
\end{equation}
with $C_A=3$, $C_f=4/3$, $D_A=8$, and the ``massive'' Mandelstam variables being now $\hat t \equiv (p_1-p_3)^2-m^2= -\hat s \tau_1$, and $\hat u \equiv (p_2-p_3)^2-m^2 = -\hat s \tau_2$. In practice, our prescription amounts to replacing the partonic transverse momentum $p_{\rm T}$ in the zero-mass expressions by the transverse mass $m_{\rm T}$ --- hence we shall name the present implementation as SACOT-$m_{\rm T}$ scheme.\footnote{This nomenclature is in the spirit of SACOT-$\chi$ scheme \cite{Tung:2001mv} in DIS, where, in essence, one replaces the Bjorken-$x$ variable (for the channels with heavy-quarks in the initial state) in the zero-mass expressions by $\chi = x\left(1+{4m^2}/{Q^2}\right)$, where $Q^2$ is the exchanged photon virtuality.}

The leading-order contribution and subtraction term for the $Q+g\rightarrow (Q \rightarrow h_3) + X$ channel are defined in a similar manner as above, so let us then discuss the contributions from light-parton fragmentation and the corresponding subtraction terms. Proceeding as in the case of initial state, we define the leading-order contribution from $g+g \rightarrow (g \rightarrow h_3) + X$ channel, originating from diagrams like (c) in Figure~\ref{fig:fig1}, by
\begin{align}
\mathrm{d}\sigma^{gg\rightarrow (g \rightarrow h_3)+X}_{\rm LO} = & \int_{z^{\rm min}}^1 \frac{\mathrm{d}z}{z} \int_{x_1^{\rm min}}^1 \mathrm{d}x_1  \, \frac{p_{\rm T}}{8\pi s^2} \frac{f_g^{h_1}(x_1,\mu^2_{\rm fact})}{x_1} \frac{f_g^{h_2}(x_2^{\rm min},\mu^2_{\rm fact})}{x_2^{\rm min}} \label{eq:gg-g} \\
& D_{g \rightarrow h_3}(z,\mu^2_{\rm frag}) \frac{|\mathcal{M}_{gg\rightarrow gg}|^2}{x_1(1-\tau_2)} \,. \nonumber
\end{align}
Together with the the perturbative expression for the gluon fragmentation function (considering that the only non-zero FF at the mass threshold is $D_{Q \rightarrow h_3}$),
\begin{equation}
D_{g \rightarrow h_3}(x,\mu^2_{\rm frag}) = \left(\frac{\alpha_s}{2\pi}\right) \log\left(\frac{\mu^2_{\rm frag}}{m^2}\right) \int_x^1 \frac{\mathrm{d}\ell}{\ell} P_{qg}\left(\frac{x}{\ell}\right) D_{Q \rightarrow h_3}(\ell,\mu^2_{\rm frag}) + \mathcal{O}(\alpha_s^2) \,, \label{eq:Dg}
\end{equation}
this defines the subtraction term
\begin{align}
S^{gg\rightarrow (g \rightarrow h_3)+X} = & \left(\frac{\alpha_s}{2\pi}\right) \log\left(\frac{\mu^2_{\rm frag}}{m^2}\right) \int_{z^{\rm min}}^1 \frac{\mathrm{d}z}{z} \int_{x_1^{\rm min}}^1 \mathrm{d}x_1  \, \frac{p_{\rm T}}{8\pi s^2} \frac{f_g^{h_1}(x_1,\mu^2_{\rm fact})}{x_1} \frac{f_g^{h_2}(x_2^{\rm min},\mu^2_{\rm fact})}{x_2^{\rm min}} \\
& \int_z^1 \frac{\mathrm{d}\ell}{\ell} P_{qg}\left(\frac{z}{\ell}\right) D_{Q \rightarrow h_3}(\ell,\mu^2_{\rm frag})
\frac{|\mathcal{M}_{gg\rightarrow gg}|^2}{x_1(1-\tau_2)} \,, \nonumber
\end{align}
where now
\begin{equation}
|\mathcal{M}_{gg\rightarrow gg}|^2 = 16\pi^2\alpha_s^2 \frac{4C_A^2}{D_A} \left[3-\tau_1 \tau_2-\frac{\tau_1}{\tau_2^2} - \frac{\tau_2}{\tau_1^2} \right] \,.
\end{equation}

The subtractions required to cancel the large logarithm originating from diagram (d) in Figure~\ref{fig:fig1} goes slightly different than the above cases. The reason is that the contributions from $g+g \rightarrow (Q \rightarrow h_3) + X$ channel (part of the inclusive heavy-quark cross sections) that we here use to determine the subtraction terms, are included using the full mass-dependence. Therefore, the subtraction term required to cancel the large logarithm that occurs when final-state heavy quark emits a collinear gluon is
\begin{align}
S^{gg\rightarrow (Q \rightarrow h_3)+X} = & \left(\frac{\alpha_s}{2\pi}\right) \log\left(\frac{\mu^2_{\rm frag}}{m^2}\right) \int_{z^{\rm min}}^1 \frac{\mathrm{d}z}{z} \int_{x_1^{\rm min}}^1 \mathrm{d}x_1  \, \frac{p_{\rm T}}{8\pi s^2} \frac{f_g^{h_1}(x_1,\mu^2_{\rm fact})}{x_1} \frac{f_g^{h_2}(x_2^{\rm min},\mu^2_{\rm fact})}{x_2^{\rm min}} \label{eq:SggQ} \\
& \int_z^1 \frac{\mathrm{d}\ell}{\ell} P_{qq}\left(\frac{z}{\ell}\right) D_{Q \rightarrow h_3}(\ell,\mu^2_{\rm frag})
\frac{|\mathcal{M}_{gg\rightarrow Q\overline{Q}}|^2}{x_1(1-\tau_2)} \,, \nonumber
\end{align}
where $P_{qq}(z)=C_f\left[\frac{1+z^2}{(1-z)_+} + \frac{3}{2}\delta(1-z)\right]$ is the quark-to-quark splitting function, and the matrix element \cite{Nason:1989zy},
\begin{align}
|\mathcal{M}_{gg\rightarrow Q\overline{Q}}|^2 & = 16\pi^2\alpha_s^2 \frac{2T_f}{D_A} \left(\tau_1^2 + \tau_2^2 +\rho - \frac{\rho^2}{4 \tau_1 \tau_2}\right) \left(\frac{C_f}{\tau_1 \tau_2} - C_A \right) \,, \\
\rho & = 4m^2/(x_1 x_2 s) \,,
\end{align}
now carries the full mass dependence. In order to recover the standard $\overline{\rm MS}$ zero-mass results at high $P_{\rm T}$ we must still compensate for the fact that the $m \rightarrow 0$ limit in the massive calculation does not exactly match that of usual massless $\overline{\rm MS}$, but some finite differences remain as a relic of a different regularization procedure. As explained in Ref.~\cite{Kniehl:2005mk}, this can be effectively achieved by replacing Eq.~(\ref{eq:SggQ}) by
\begin{align}
S^{gg\rightarrow (Q \rightarrow h_3)+X} = & \int_{z^{\rm min}}^1 \frac{\mathrm{d}z}{z} \int_{x_1^{\rm min}}^1 \mathrm{d}x_1  \, \frac{p_{\rm T}}{8\pi s^2} \frac{f_g^{h_1}(x_1,\mu^2_{\rm fact})}{x_1} \frac{f_g^{h_2}(x_2^{\rm min},\mu^2_{\rm fact})}{x_2^{\rm min}} \label{eq:SggQnew} \\
& \int_z^1 \frac{\mathrm{d}\ell}{\ell} d_{QQ}\left(\frac{z}{\ell}\right) D_{Q \rightarrow h_3}(\ell,\mu^2_{\rm frag})
\frac{|\mathcal{M}_{gg\rightarrow Q\overline{Q}}|^2}{x_1(1-\tau_2)} \,, \nonumber 
\end{align}
where $d_{QQ}$ is the partonic fragmentation function \cite{Mele:1990cw,Melnikov:2004bm},
\begin{equation}
d_{QQ}(z) \equiv \left(\frac{\alpha_s}{2\pi}\right) C_f \left\{\frac{1+z^2}{1-z} \left[\log\left(\frac{\mu^2_{\rm frag}}{m^2}\right) -2\log(1-z)-1 \right] \right\}_+ \,.
\end{equation}
Also the renormalization procedure applied in FFNS calculations is slightly different than in the purely zero-mass case. Indeed, the FFNS results of Ref.~\cite{Nason:1989zy} are obtained in a so-called decoupling scheme, where $\alpha_s$ runs with only light partons (gluons + 3 light-flavour quarks). Above the transition scale $Q_{\rm t}$, also the heavy-quark is considered as being ``active'' in the running of $\alpha_s$, and the matching between the two schemes induces an additional contribution. Specifically, we must add a term
\begin{equation}
 -\alpha_s \frac{2T_f}{3\pi}\log \left( \frac{\mu^2_{\rm ren}}{\mu^2_{\rm fact}} \right) \mathrm{d}\sigma^{gg\rightarrow (Q \rightarrow h_3)+X}_{\rm LO}, \label{eq:ren1}
\end{equation}
as explained in Ref.~\cite{Cacciari:1998it}. 

The same line of reasoning is applied when defining the subtraction terms for $q+\overline{q} \rightarrow Q + X$ and $q+g \rightarrow Q + X$ channels and the emerging leading-order contributions from $q+\overline{q} \rightarrow (g \rightarrow h_3) + X$, $q+Q \rightarrow (Q \rightarrow h_3) + X$, and $q+g \rightarrow (g \rightarrow h_3) + X$ channels. As in Eq.~(\ref{eq:SggQnew}), the definition of the $S^{q\overline{q}\rightarrow (Q \rightarrow h_3) + X}$ subtraction term involves the partonic fragmentation function $d_{QQ}$, and a term 
\begin{equation}
 -\alpha_s \frac{2T_f}{3\pi}\log \left( \frac{\mu^2_{\rm ren}}{m^2} \right) \mathrm{d}\sigma^{q\overline{q}\rightarrow (Q \rightarrow h_3)+X}_{\rm LO}, \label{eq:ren2}
\end{equation}
is added to recover the $\overline{\rm MS}$ renormalization scheme \cite{Cacciari:1998it}. In addition, our full results include the contributions from all other partonic subprocesses whose inclusion does not require a preparation of subtraction terms at the perturbative order we work at. The NLO $\mathcal{O}(\alpha_s^3)$ contributions, taken from Ref.~\cite{Aversa:1988vb} are included as well. We stress that when including these terms in our SACOT-$m_{\rm T}$ scheme, we consistently retain the kinematics which they inherit from $Q\overline{Q}$ pair-creation process as explained earlier. In practice this is done by trading the massless variables $v$ and $w$ used in Ref.~\cite{Aversa:1988vb} by their massive counterparts $v \rightarrow 1-\tau_1$ and $w \rightarrow \tau_2/(1-\tau_1)$, see e.g. Sect.~2 of Ref.~\cite{Kniehl:2005mk}, and imposing the proper integration limits explained in Section~\ref{sec:Generalstructureandkinematics}. In this way, we already implicitly define the subtraction terms that would be required at a next-to-NLO (NNLO) -level calculation. 

In comparison to the earlier works \cite{Kniehl:2004fy,Kniehl:2005mk}, the most notable advantage of the SACOT-$m_{\rm T}$ scheme is that the cross sections remain finite in the $P_{\rm T} \rightarrow 0$ limit. Indeed, in Refs.~\cite{Kniehl:2004fy,Kniehl:2005mk} at least part of the contributions not coming directly from flavour-creation processes are included using purely zero-mass formalism, and give rise to a divergent $P_{\rm T}^{-n}$ behaviour at $P_{\rm T} \rightarrow 0$ limit. The difficulty will not be completely resolved at NNLO either, though the divergences may be a bit ``softer''. In Ref.~\cite{Kniehl:2015fla} these divergent contributions were excluded at small $P_{\rm T} = 0$ by maintaining the factorization and fragmentation scale at (or below) the heavy-quark mass threshold until large-enough $P_{\rm T}$. This procedure leads to finite cross sections in the $P_{\rm T} \rightarrow 0$ limit, but causes certain unphysical slope change near the $P_{\rm T}$ value where the factorization and fragmentation scales go above the mass threshold --- we will come back to this in Section~\ref{sec:ComparisonwithLHCbdata} (see also see Fig.~6 in Ref.~\cite{Kniehl:2015fla}). In our case -- and this applies also for the fixed-order calculations -- the divergent behaviour is regulated by the heavy-quark mass and leads to finite cross sections even at $P_{\rm T} = 0$ (at any perturbative order) without a need to fine tune the scale choices. Technically, this happens because the lower limits for the scaling variables $\tau_1$ and $\tau_2$ appearing in the squared matrix elements are not zero but limited by the heavy-quark mass.

\subsection{Numerical implementation}
\label{sec:Numericalimplementation}

Our numerical realization of the GM-VFNS scheme described above is crafted around the public \textsc{INCNLO} \cite{Aversa:1988vb,INCNLO} and Mangano-Nason-Ridolfi (MNR) \cite{Mangano:1991jk,MNR} codes. The former provides the zero-mass matrix elements, and the latter one the one-particle inclusive heavy-quark cross section of Ref.~\cite{Nason:1989zy}. As already noted in Ref.~\cite{dEnterria:2013sgr}, in order to obtain reliable numerical results from \textsc{INCNLO} at high $\sqrt{s}$ away from the midrapidity $|y|\gg 0$, the numerical stability of the original code has had to be improved, see p.30-32 in Ref.~\cite{Helenius:2014jza} for a detailed explanation. Schematically, we compute
\begin{align}
 & \left[
 \sum_{ijk} \sigma^{ij\rightarrow (k \rightarrow h_3)+X} - 
 \left(
 \sigma^{gg\rightarrow ({Q} \rightarrow h_3)+X} +
 \sigma^{q\overline{q}\rightarrow ({Q} \rightarrow h_3)+X} +
 \sigma^{qg\rightarrow ({Q} \rightarrow h_3)+X}
 \right)
 \right]_{m=0} \label{eq:schema} \\ 
 & \hspace{3.2cm} + \left(
 \sigma^{gg\rightarrow ({Q} \rightarrow h_3)+X} +
 \sigma^{q\overline{q}\rightarrow ({Q} \rightarrow h_3)+X} +
 \sigma^{qg\rightarrow ({Q} \rightarrow h_3)+X}
 \right)_{m \neq 0} \nonumber \\
 & \hspace{3.2cm} - {\rm subtraction \ terms} \nonumber \,, 
\end{align}
where the inclusion of charge-conjugate contributions and shuffling between the initial-state partons is implicit. That is, from the full zero-mass result we subtract the zero-mass contributions of $g+g \rightarrow Q + X, \ q+\overline{q} \rightarrow Q + X, \ {\rm and} \ q+g \rightarrow Q + X$ channels which we add back using the full mass dependence. The subtraction terms provide the proper matching. Towards high $P_{\rm T}$, only the first sum term in Eq.~(\ref{eq:schema}) survives --- others add up to zero. In the numerical evaluation we have used NNPDF31\_nlo\_pch\_as\_0118 variable-flavour-number PDFs and the corresponding running strong coupling $\alpha_s$ \cite{Ball:2017nwa}. This is the latest NNPDF fit assuming no intrinsic charm content in the proton. The PDFs are interfaced by using LHAPDF 6 library \cite{Buckley:2014ana}. The introduced framework is applicable to production of any hadrons involving heavy quarks but in this work we consider only D-meson production due to good availability and precision of the experimental data from LHC experiments. In particular, we will focus on D$^0$-meson production, with the data from LHCb \cite{Aaij:2016jht,Aaij:2013mga,Aaij:2015bpa}, ALICE \cite{Adam:2016ich}, and CMS \cite{Sirunyan:2017xss} (though not yet available) extending to small $P_{\mathrm{T}}$ region, which is where our SACOT-$m_{\mathrm{T}}$ scheme mostly differs from other GM-VFNS implementations
(the LHCb collaboration has also measured D$^\pm$ at small $P_{\mathrm{T}}$). For  D$^*$ mesons there would be more recent FF analyses available \cite{Anderle:2017cgl, Soleymaninia:2017xhc} but for D$^0$ we use KKKS08 \cite{Kneesch:2007ey} FFs, which is the only available FF set for D$^0$'s. For the charm-quark mass we use $m_{\rm charm} = 1.51 \, {\rm GeV}$ in accordance with the used PDF set. The input charm mass in KKKS08 analysis was $1.50 \, {\rm GeV}$ so the pairing with NNPDF3.1 is consistent. Our default scale choice will be $\mu_{\rm ren} = \mu_{\rm fact} = \mu_{\rm frag} = \sqrt{P_{\rm T}^2 + m_{\rm charm}^2} \,$, and for the D-meson mass we use $M=1.87 \, {\rm GeV}$. The small contribution from b-quark fragmentation is retained in the calculation neglecting the finite b-quark mass. The D mesons from B-meson decays have been excluded from the LHCb and ALICE data we discuss later on, but as the KKKS08 FFs include these feed-down D mesons as well, there is no fully consistent way to exclude them without explicitly evaluating the D$^0$ meson spectra from B-meson decays and subtracting it from the fully inclusive cross section. However, the contributions from B decays are very small, less than 1\% in the integrated inclusive D$^0$-meson cross section of ALICE \cite{Adam:2016ich}.

In order to compare with another popular approach, we have used here the \textsc{Powheg} method \cite{Frixione:2007nw} in which the $Q\overline{Q}$ production at FFNS is matched with the \textsc{Pythia} parton shower providing NLO accuracy for the matrix element generation and leading-log resummation from the parton shower. In practice, we have first generated $\mathrm{c}\overline{\mathrm{c}}$ events with the \textsc{hvq} part \cite{Frixione:2007vw} of \textsc{Powheg-Box} generator \cite{Alioli:2010xd}. The generated events are then fed into \textsc{Pythia} (version 8.230) \cite{Sjostrand:2014zea} which generates the $p_{\mathrm{T}}$-ordered parton shower and hadronizes the events using the implemented Lund string model with parameter values from the default \textsc{Monash} tune \cite{Skands:2014pea}. The D$^0$ mesons (and its charge conjugate) are then picked up from the hadronized final state and binned in $Y$ and $P_{\mathrm{T}}$. The same NNPDF3.1 PDFs as for the GM-VFNS calculations have been used for the event generation in \textsc{Powheg} and also in showering within \textsc{Pythia}. The sensitivity of the \textsc{Pythia} shower to PDFs is very mild as they affect only the initial-state emission probabilities and there only ratios of PDFs are involved. In \textsc{Powheg} generation the default scale choice is $\mu_{\rm ren} = \mu_{\rm fact} = \sqrt{p_{\rm T}^2 + m_{\rm charm}^2} \,$ with $m_{\rm charm} = 1.5\,{\rm GeV}$. We have not explicitly introduced the matching terms, Eqs.~(\ref{eq:ren1}) and (\ref{eq:ren2}), at the heavy-quark mass thresholds as their effect has been found small in the $P_{\rm T}$ range of LHCb data \cite{Gauld:2015yia}. Indeed, with $\mu_{\rm ren} = \mu_{\rm fact}$ the first matching term in Eq.~(\ref{eq:ren1}) is zero, and the second term in Eq.~(\ref{eq:ren1}) is small as the LO contribution of $q\overline{q}$ channels is small. As discussed in Ref.~\cite{Gauld:2015yia}, \textsc{Powheg}+\textsc{Pythia} yields very similar results as e.g. FONLL \cite{Cacciari:1998it,Cacciari:2001td} or Madgraph5\_aMC@NLO \cite{Frixione:2002ik,Alwall:2014hca} approaches in the kinematic domain of LHCb.

\section{Results}
\label{sec:Results}

In this section, we will first illustrate some features of our calculation that we have studied numerically and then compare with the available experimental LHC data.

\subsection{Consistency checks and other trivia}
\label{sec:Consistencychecks}

\begin{figure}[htb!]
\centering
\includegraphics[width=0.49\textwidth]{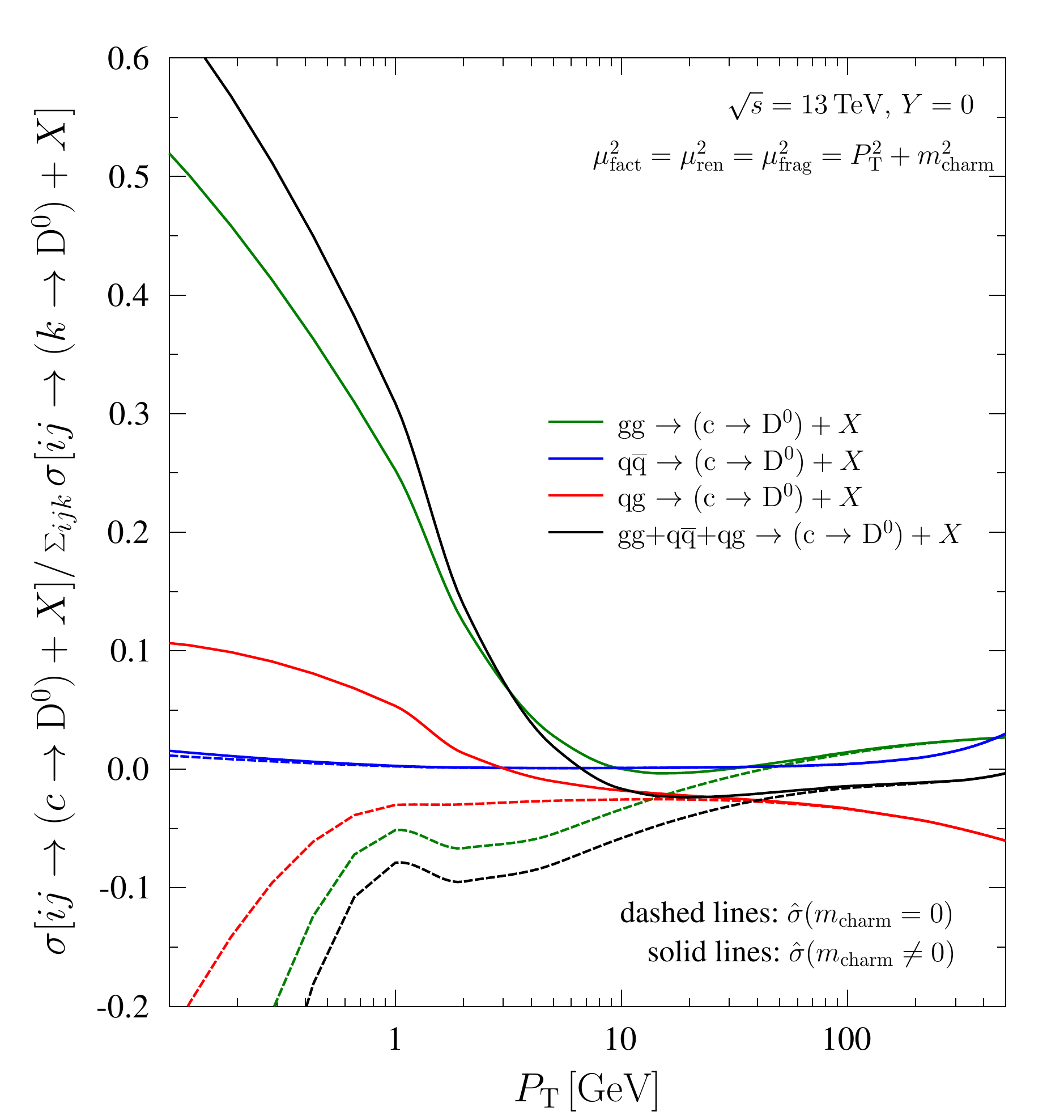} 
\includegraphics[width=0.50\textwidth]{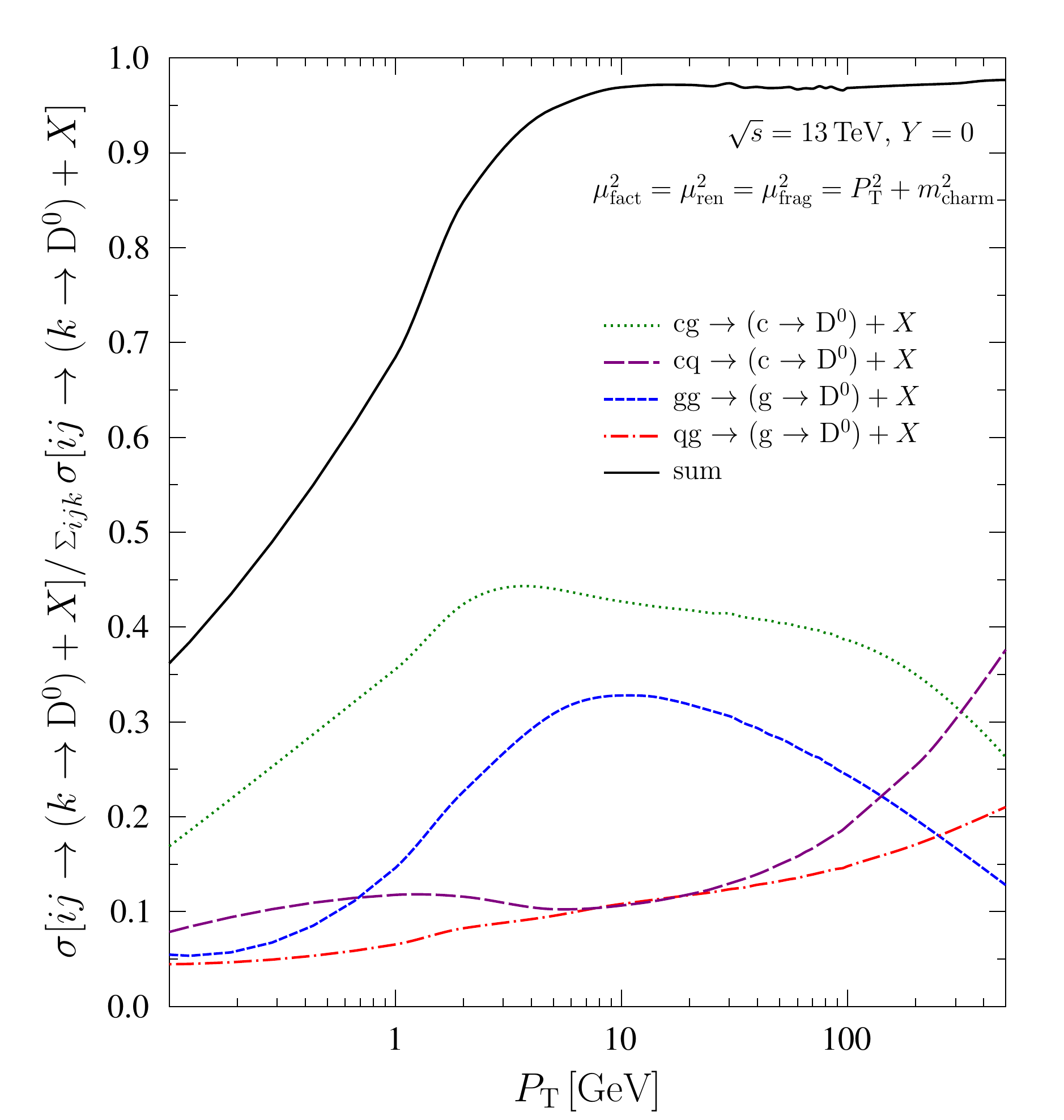} 
\caption{{\bf Left-hand panel:} Contributions of light-parton-to-heavy-quark processes normalized to the full result including all the channels. The solid curves correspond to the complete GM-VFNS calculation and the dashed curves are from a calculation with zero-mass matrix elements. Contributions from the $\overline{c}$ fragmentation and shuffling between the initial-state partons is implicit. {\bf Right-hand panel:} Contributions from channels with heavy quark in the initial state or where the D meson is produced from gluon fragmentation. Results are normalized to the full result including all the channels. Contributions in which $c$ is replaced by $\overline{c}$, and interchange between the initial-state partons is implicit. 
} 
\label{fig:check}
\end{figure}

We begin to fold out the numerical results by showing in the left-hand panel of Figure~\ref{fig:check} contributions from the channels where the $\mathrm{c\overline{c}}$ pair is explicitly produced. Here, we have taken $\sqrt{s}=13\,{\rm TeV}$ and $Y=0$. The solid curves are from the calculation with full mass dependence including the relevant subtraction terms and the dashed ones correspond to the evaluation with zero-mass Wilson coefficients (but still retaining the $Q\overline{Q}$ kinematics). The results are normalized by the full GM-VFNS calculation including all the partonic channels. At high $P_{\rm T}$ the solid and dashed curves merge which provides a non-trivial, strong check on the consistency of our implementation. Towards $P_{\rm T} \rightarrow 0$ the two sets of curves, however, behave completely differently: Whereas all channels of the ``massive'' calculation yield a positive contribution at $P_{\rm T} \rightarrow 0$ limit, even the overall result with zero-mass matrix elements remains negative.

As can be seen from the left-hand panel of Figure~\ref{fig:check}, the overall contribution from the channels where the $\mathrm{c\overline{c}}$ pair is explicitly produced, is only a few percents from $P_{\rm T} \sim 5 \, {\rm GeV}$ onwards. In fact, almost the entire cross sections in this region accumulates from the partonic subprocesses with heavy quarks in the initial state or gluon fragmentation, around 50\% coming from each of these two sources. This is demonstrated in the right-hand panel of Figure~\ref{fig:check} where we plot the contributions from these channels, normalized to the full GM-VFNS result. The balance between the contributions shown in the left- and right-hand panels of Figure~\ref{fig:check} depends rather strongly on the scale choices at low $P_{\rm T}$, and the pace at which the contributions in the right-hand panel begin to dominate can be controlled by adjusting the scales. Indeed, using a lower scale than our default choice, the contributions shown in the right-hand panel would begin to dominate at higher $P_{\rm T}$ than now shown in Figure~\ref{fig:check}. As already mentioned in Section~\ref{sec:Introduction}, it was exactly this property that was taken advantage of in Refs.~\cite{Kniehl:2015fla,Benzke:2017yjn} to suppress the divergent contributions at low $P_{\rm T}$.

\begin{figure}[htb!]
\centering
\includegraphics[width=0.49\textwidth]{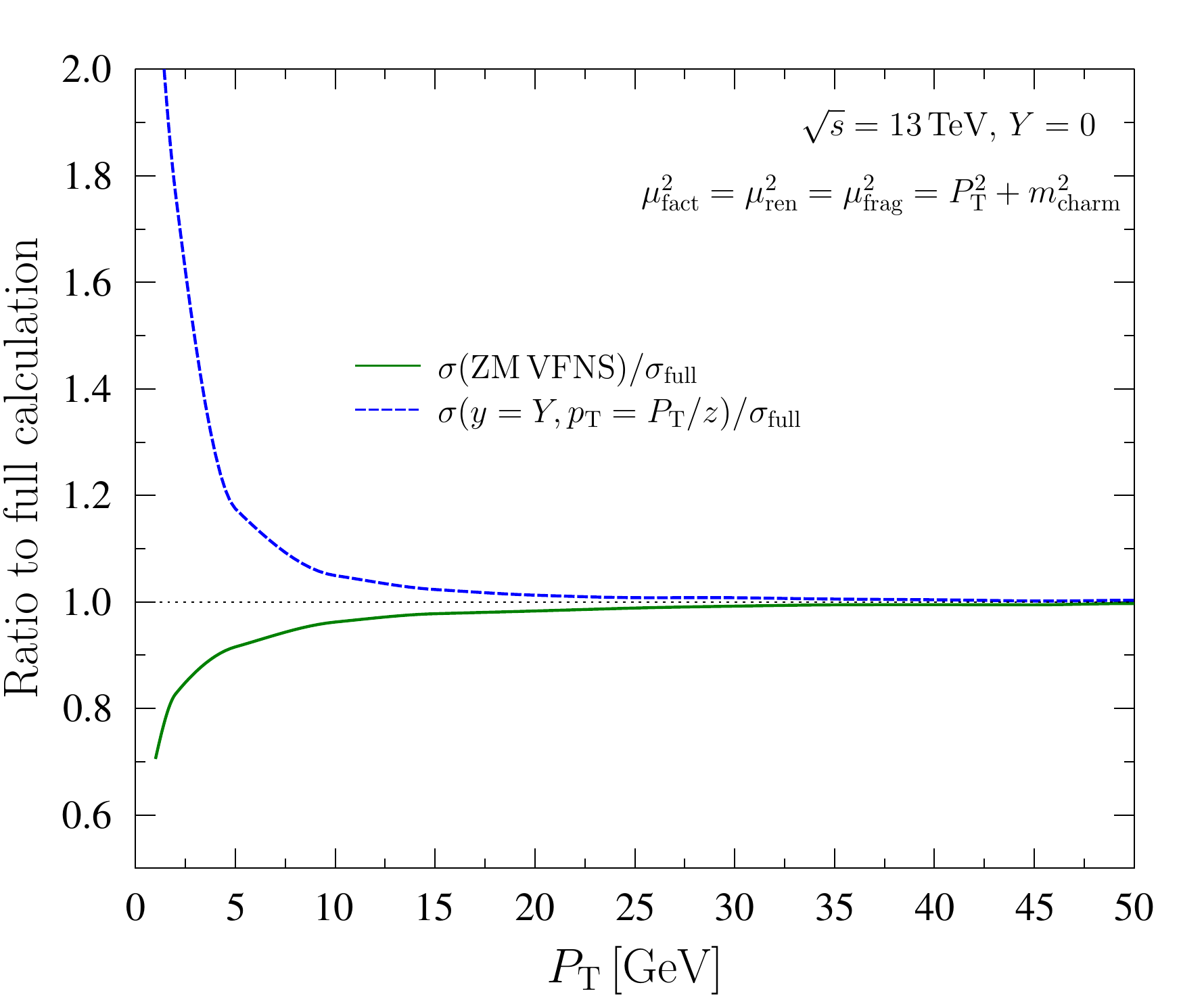}
\includegraphics[width=0.49\textwidth]{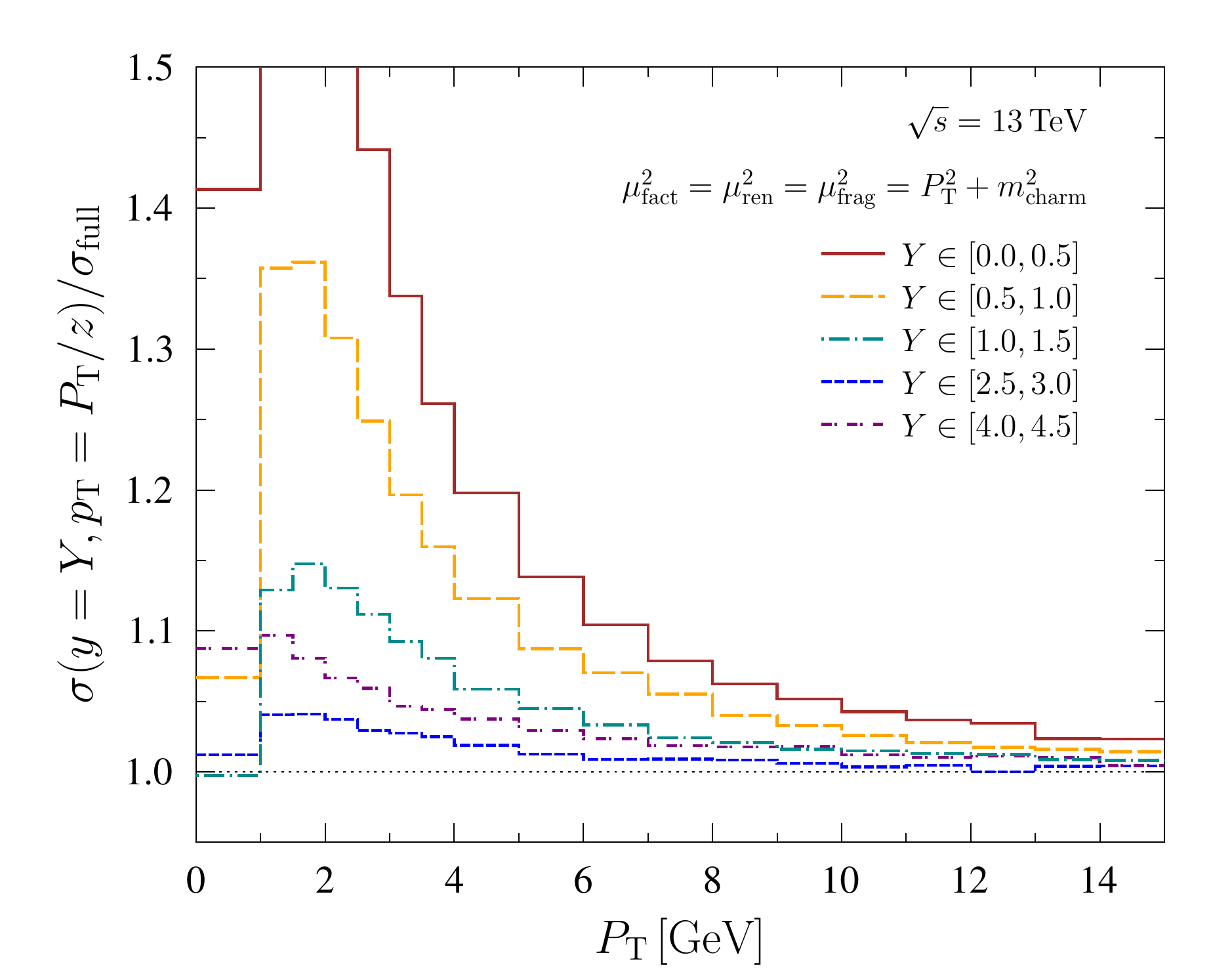}
\caption{
{\bf Left-hand panel:} A calculation at $Y=0$ taking $m=0$ in matrix elements (green curve), and the calculation ignoring the mass dependencies in the fragmentation variable $z$ (blue dashed curve). In both cases, the results are normalized to the full calculation. {\bf Right-hand panel:} Relative effect also at $Y > 0$ when ignoring the mass dependencies in the fragmentation variable $z$.
} 
\label{fig:systeff}
\end{figure}

In the left-hand panel of Figure~\ref{fig:systeff} we estimate the effects of charm-quark and D-meson masses in our cross sections at $Y=0$. The green curve corresponds to a ZM-VFNS calculation (but still using the aforementioned default scale choice) normalized with the full GM-VFNS result. In accord with what was seen in Figure~\ref{fig:check}, we observe that neglecting the charm mass leads to a lower cross section at low $P_{\rm T}$ due to increasingly negative contributions from $g+g \rightarrow (\mathrm{c} \rightarrow h_3) + X$ and $g+q \rightarrow (\mathrm{c} \rightarrow h_3) + X$ channels in ZM-VFNS. The blue dashed curve corresponds to putting $m,M=0$ in Eq.~(\ref{eq:pt_PT}) and Eq.~(\ref{eq:y_Y}), that is, ignoring the mass dependence in the fragmentation variable $z$. We observe that this manoeuvre leads to increased cross sections. The origin of the effect can be understood relatively easily on the grounds of Eqs.~(\ref{eq:pt_PT}) and (\ref{eq:y_Y}) from which it follows that
\begin{equation}
p_{\rm T} \geq P_{\rm T}/z, \quad |y| \geq |Y| \,, \label{eq:partonic_vs_hadronic_y_pT}
\end{equation}
when $m<M$. That is, for fixed $Y$ and $P_{\rm T}$ the partonic cross sections are probed at larger $y$ and larger $p_{\rm T}$ in comparison to the massless kinematics. Since the partonic cross sections decrease steeply, particularly with increasing $p_{\rm T}$, also the hadronic cross sections are consequently lower. In our framework, this explains why the hadronic cross sections are suppressed in the presence of non-zero masses. This is in contrast to what has been found in Ref.~\cite{Kniehl:2015fla} in the case of B mesons, though there a different version of the fragmentation variable $z$ was used. Moreover, in Ref.~\cite{Albino:2008fy} a very similar definition of $z$ as in Ref.~\cite{Kniehl:2015fla} was adopted and there, in turn, the mass effects led to suppressed cross sections (as in our case). To clear up the systematics of different definitions of the fragmentation variable warrants a separate study which is beyond our scope here. Nevertheless, the effects of finite hadron and quark masses can be non-negligible up to $P_{\rm T} \sim 20 \, {\rm GeV}$ which signifies a possibly considerable source of theoretical uncertainty, given that the definition of fragmentation variable $z$ is ambiguous. In the context of the present definition of $z$ these effects will, however, get milder towards larger $Y$. This can be easily understood from Eq.~(\ref{eq:pt_PT}) from which it follows that $p_{\rm T} \sim P_{\rm T}/z$ when $Y \gg 0$. Thus, with the present definition of the fragmentation variable $z$, part of the significant effect found at $Y=0$ can be expected to melt away. This is demonstrated in the right-hand panel of Figure~\ref{fig:systeff} where we show the impact of massive fragmentation variable also in the forward direction. The differences between massless and massive fragmentation variable get clearly suppressed when moving to larger $Y$. At very large $Y$ the effect starts to rise again as the partonic $y$ spectrum gets steeper (near $y \sim 0$ it is quite flat) and the condition $|y| \geq |Y|$ of Eq.~(\ref{eq:partonic_vs_hadronic_y_pT}) begins to matter increasingly. 

\begin{figure}[htb!]
\centering
\includegraphics[width=0.49\textwidth]{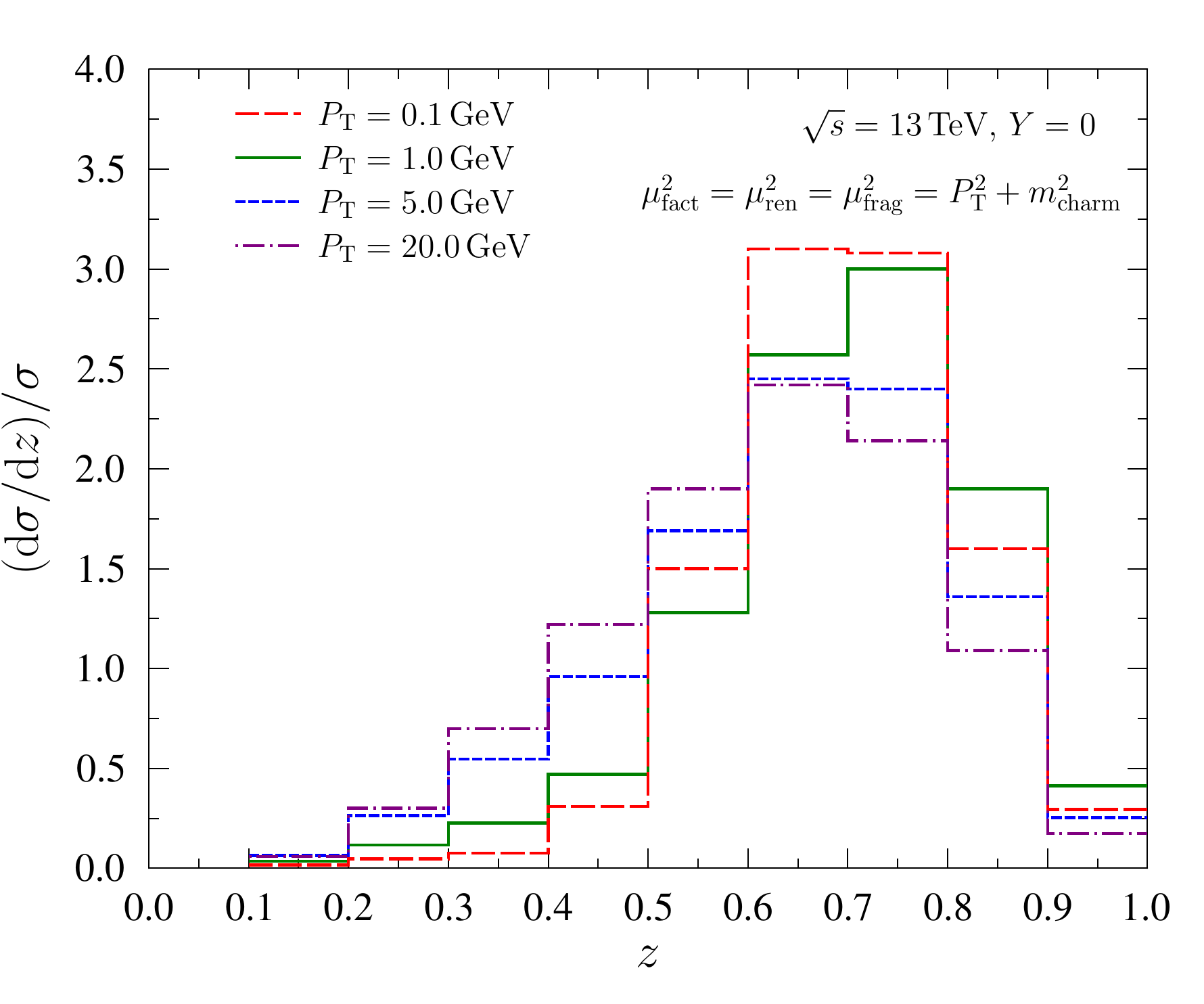}
\caption{
The $z$ distributions at $\sqrt{s}=13\,{\rm TeV}$ and $Y=0$ for four fixed values of $P_{\rm T}$.
} 
\label{fig:z_distributions}
\end{figure}

\begin{figure}[htb!]
\centering
\includegraphics[width=1.00\textwidth]{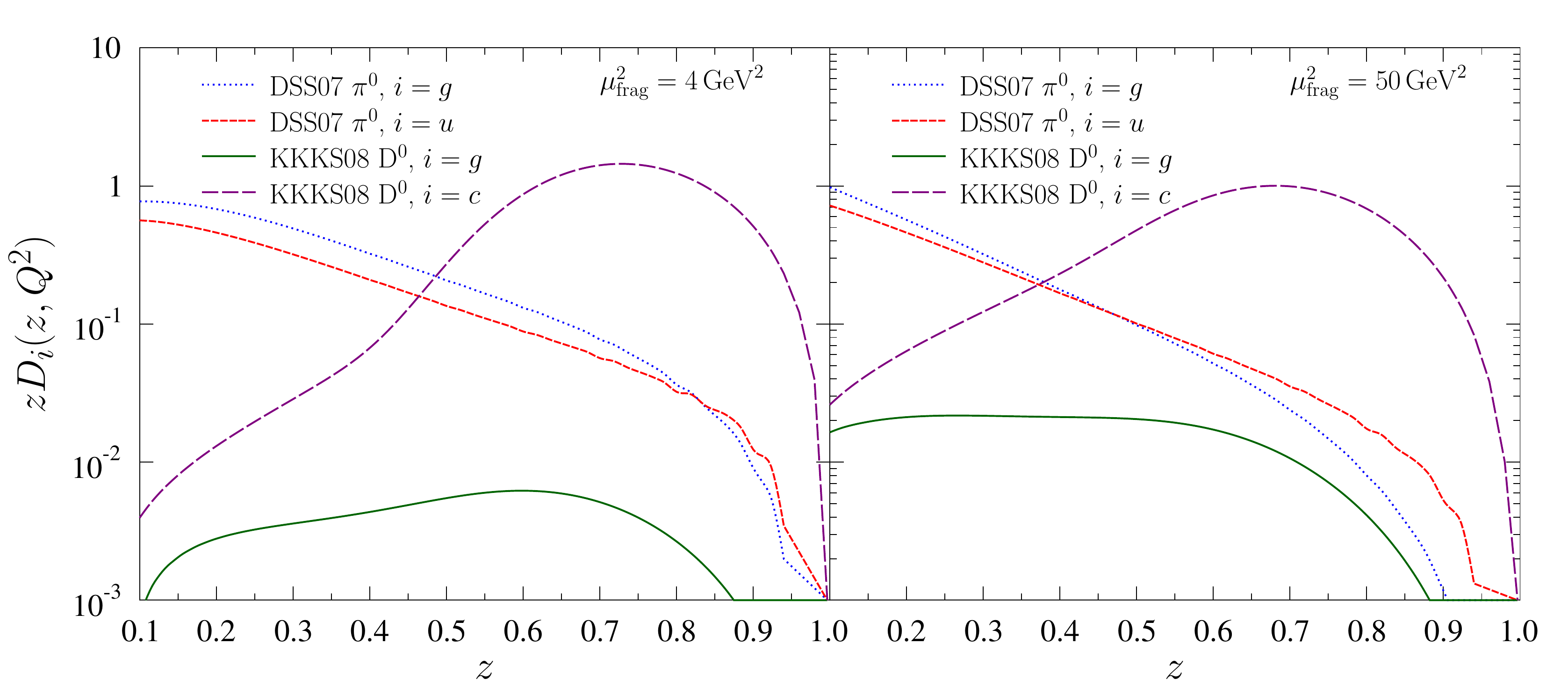}
\caption{{\bf Left-hand panel:} A comparison between $\pi^0$ FFs from the DSS07 analysis \cite{deFlorian:2007aj} and D$^0$ FFs from the KKKS08 fit \cite{Kneesch:2007ey} at $\mu_{\rm frag}^2=4\,{\rm GeV}^2$. {\bf Right-hand panel:} As the left-hand panel but at $\mu_{\rm frag}^2=50\,{\rm GeV}^2$.} 
\label{fig:DSS_vs_KKKS}
\end{figure}

The validity of our calculation towards low $P_{\rm T} \rightarrow 0$ could be potentially compromised by the unstable fixed-order NLO scale evolution of the fragmentation functions below $z \sim 0.1$, stemming from singular $\alpha_s^2(\log^2 z)/z$ terms in the time-like NLO quark-to-gluon and gluon-to-gluon splitting functions, see e.g. Ref.~\cite{deFlorian:1997zj}. The proper treatment of this region requires resummation in both, splitting functions and Wilson coefficients \cite{Anderle:2016czy}. To exclude contributions from the unstable region we have imposed a condition $z>0.1$ when computing the cross sections. When doing so we must then make sure that this cut is not overly strict, i.e. that the contribution outside of the introduced cut is negligible. The reason why the $P_{\rm T} \rightarrow 0$ limit could pose a problem, can be easily understood: As discussed e.g. in Refs.~\cite{Eskola:2002kv,dEnterria:2013sgr}, approximating the convolution of partonic cross sections and PDF by a power law,
\begin{equation}
\sum _{ijk} \int \mathrm{d}x_1 \mathrm{d}x_2 f_i^{h_1}(x_1,\mu^2_{\rm fact}) f_j^{h_2}(x_2,\mu^2_{\rm fact}) \frac{\mathrm{d}\hat{\sigma}^{ij\rightarrow k}(\tau_1, \tau_2, \mu^2_{\rm ren}, \mu^2_{\rm fact}, \mu^2_{\rm frag})}{\mathrm{d}p_{\rm T} \mathrm{d}y} \approx \frac{C_k}{p_{\rm T}^{n}}  \,, \label{eq:powerlaw}
\end{equation}
where $n>0$ and $C_k$ does not depend on $p_{\rm T}$, one gets
\begin{equation}
 \frac{\mathrm{d}\sigma(h_1 + h_2 \rightarrow h_3 + X)}{\mathrm{d}P_{\rm T}\mathrm{d}Y\mathrm{d}z} \approx 
P_{\rm T}^{-n} \sum_k {C_k} \left[z^{n-1} D_{l \rightarrow h_3}(z,\mu^2_{\rm frag}) \right] \,, \label{eq:zsupress}
\end{equation}
in the zero-mass approximation. If the partonic spectrum drops sufficiently strongly in $p_{\rm T}$ (i.e. the exponent $n$ is large enough), the factor $z^{n-1}$ efficiently eliminates the contributions from the problematic low-$z$ domain. However, in the low-$P_{\rm T}$ region the LHC data \cite{Aaij:2013mga,Aaij:2015bpa,Aaij:2016jht,Aaij:2017gcy,Adam:2016ich} show that the hadronic D-meson cross sections tend to level off towards $P_{\rm T} \rightarrow 0$, see Figure~\ref{fig:LHCb13} ahead. That is, the exponent $n$ in Eq.~(\ref{eq:powerlaw}) decreases and the mechanism above is not as effective in suppressing the small-$z$ contributions. In Figure~\ref{fig:z_distributions} we show $z$ distributions obtained directly from the full calculation for a few fixed values of $P_{\rm T}$. Unlike could have been expected on the basis of the above discussion, the cross sections are found practically inert to the small $z$ region even at very small $P_{\rm T}$. Here, the explanation seems to be in the form of the D-meson fragmentation functions which at low $\mu_{\rm frag}$ are clearly suppressed in the small-$z$ region as shown in Figure~\ref{fig:DSS_vs_KKKS}. Towards higher $\mu_{\rm frag}$ the small-$z$ tails go up but then also the probed $p_{\rm T}$ is larger (larger exponent $n$) and the contributions are suppressed by virtue of Eq.~(\ref{eq:zsupress}). The behaviour of the D-meson fragmentation functions are quite different in comparison to e.g. typical pion fragmentation function as demonstrated in Figure~\ref{fig:DSS_vs_KKKS} as well. All in all, the cross sections get hardly any contributions from the small $z$ region. At NNLO and beyond, also the PDF evolution becomes similarly unstable at small $x$, and resummation \cite{Bonvini:2017ogt} appears to be required in order to optimally reproduce the small-$x$ HERA data at low $Q^2$ \cite{Ball:2017otu,Abdolmaleki:2018jln}. However, at the NLO level we work at, these issues are not yet that pressing. Indeed, based on Ref.~\cite{Ball:2017otu}, the effects of resummation in PDFs are only modest at NLO, and thus we expect that the small-$x$ resummation would lead to only subleading effects in comparison to the very large scale uncertainty in low $P_{\rm T}$ D-meson production, see e.g. Figure~\ref{fig:LHCb13} ahead.

\begin{figure}[htb!]
\centering
\includegraphics[width=0.490\textwidth]{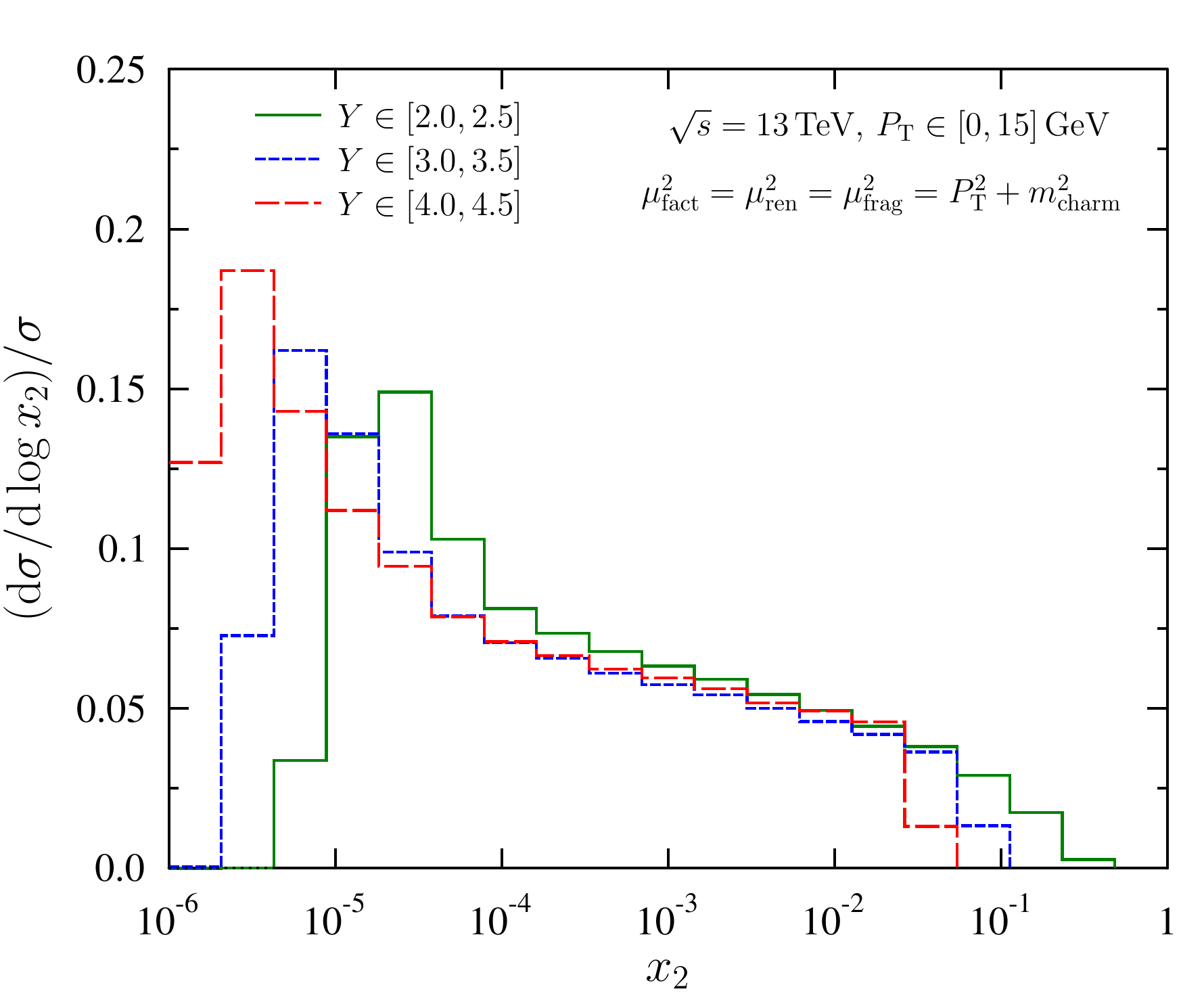}
\includegraphics[width=0.490\textwidth]{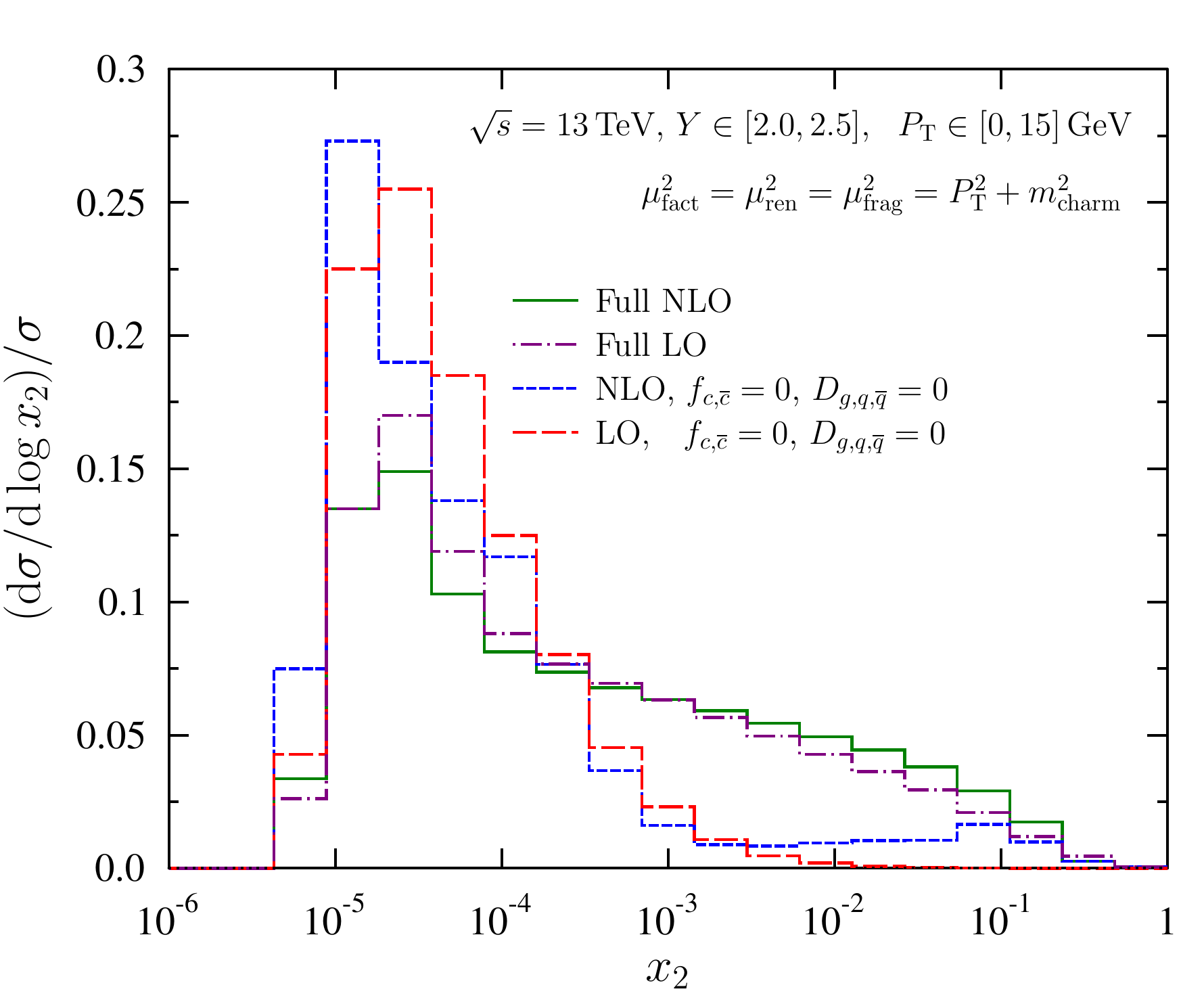}
\includegraphics[width=0.490\textwidth]{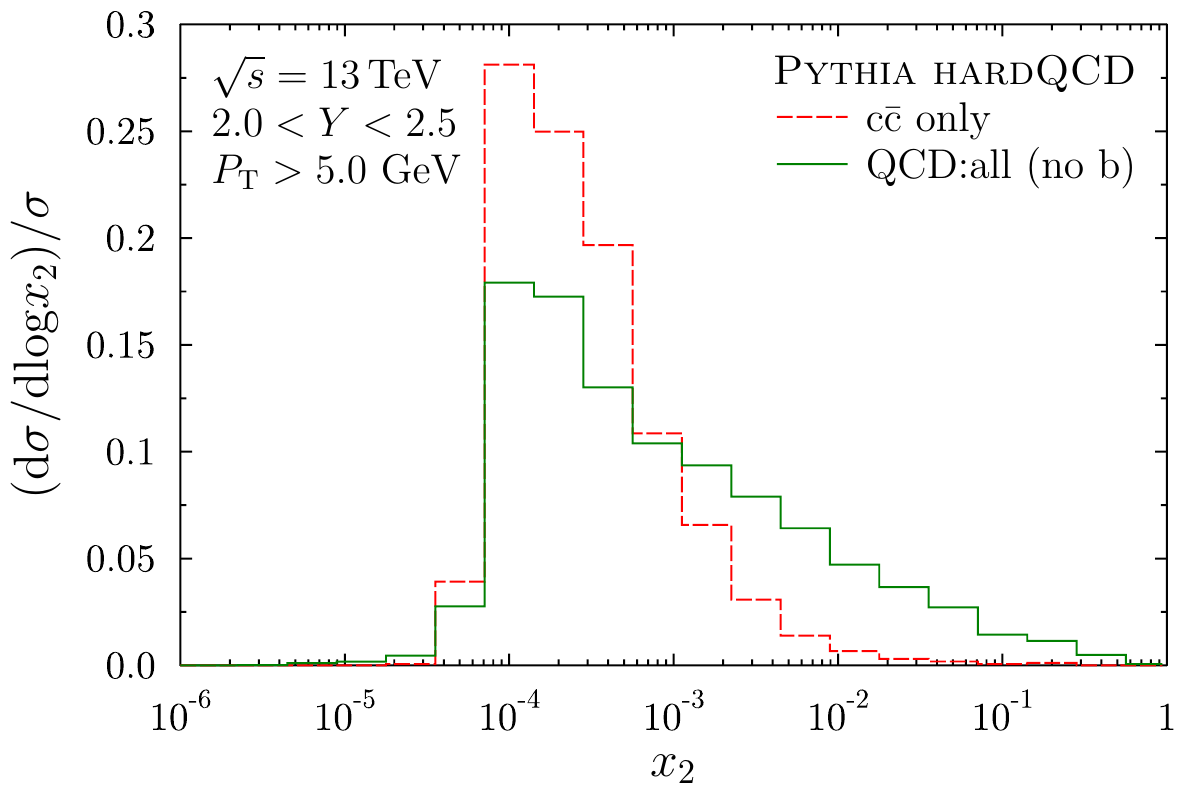}
\caption{{\bf Upper left panel:} The $x_2$ distributions for D$^0$-meson production in SACOT-$m_{\rm T}$ scheme at $\sqrt{s}=13\,{\rm TeV}$ for $Y\in[2,2.5]$ (green solid line), $Y\in[3,3.5]$ (blue dashed line), and $Y\in[4,4.5]$ (red dashed line). {\bf Upper right panel:} The $x_2$ distributions for rapidity interval $Y\in[2,2.5]$ from full NLO calculation (green solid line), full LO calculation (purple dashed-dotted line), partial NLO calculation with heavy-quark PDFs and light-parton FFs set to zero (blue dashed line), and partial LO calculation with heavy-quark PDFs and light-parton FFs set to zero (red long-dashed line). {\bf Lower panel:} The $x_2$ distributions from a \textsc{Pythia} simulation from $\mathrm{c}\overline{\mathrm{c}}$ events only (green solid line), and including all (excluding b quarks) QCD processes (red dashed line).} 
\label{fig:xdists}
\end{figure}

Before comparing with the data, we wish to shortly discuss the predominant $x$ ranges sampled by D meson production, in particular the small-$x$ sensitivity within the LHCb acceptance. To this end, the upper left-hand panel of Figure~\ref{fig:xdists} presents examples of $x_2$ distributions as obtained from our GM-VFNS calculation. The distributions are presented for $\sqrt{s}=13\,{\rm TeV}$ with typical LHCb kinematics. While the lower limits for the $x_2$ distributions are always deep in the small-$x$ domain, the distributions carry a long tail towards large $x$ --- in all of the considered cases there is a clearly non-negligible contribution coming even from the $x_2\gtrsim 10^{-2}$ region. In part, this tail originates from the NLO contributions in processes where the $\mathrm{c}\overline{\mathrm{c}}$ pair is explicitly produced, but mostly it comes from the new partonic channels that ``open'' as a result of resumming the collinear logarithms into non-zero heavy-quark PDFs and light-parton FFs. To corroborate this point, the upper right-hand panel of Figure~\ref{fig:xdists} shows various $x_2$ distributions for the bin $Y\in[2,2.5]$. The LO calculation with the heavy-quark PDFs and light-parton FFs turned off (only ``direct'' $\mathrm{c}\overline{\mathrm{c}}$) gets almost no contribution from large $x$, but when the NLO contributions are switched on, a smallish large-$x$ tail develops.\footnote{We have also verified this hierarchy with the FONLL code at parton level.} The normalized $x_2$ spectra from the full LO and NLO calculations are mutually very similar, both getting a significant contribution from $x_2 \gtrsim 10^{-3}$ unlike the contributions from ``direct'' $\mathrm{c}\overline{\mathrm{c}}$ production processes. This shows that the large-$x$ tail mostly originates from other than explicit $\mathrm{c}\overline{\mathrm{c}}$ processes. To illustrate this point even further from a different viewpoint we show, in the lower panel of Figure~\ref{fig:xdists}, also predictions from pure LO \textsc{Pythia} simulations. On one hand, when the origin of the D meson is restricted to underlying $\mathrm{c}\overline{\mathrm{c}}$ events, the $x_2$ distributions are clearly suppressed at large $x$. On the other hand, when the D mesons are allowed to be created from all partonic QCD processes (here we omitted b quarks), the large-$x$ tail emerges. All in all, the D-meson production at forward rapidity is sensitive to the PDFs at small-$x$, but the role of large-$x$ contributions is still clearly non negligible. The importance of the large-$x$ part is something that has maybe been a bit underrated in many recent articles \cite{Gauld:2015yia,Cacciari:2015fta,Zenaiev:2015rfa}, in part because their importance does not show up in fixed-order-based calculations. We note that a very similar large-$x$ behaviour is present also in isolated photon production \cite{Helenius:2014qla}. In inclusive pion production the large-$x$ tail is even more pronounced \cite{Helenius:2014qla} due to different behaviour of parton-to-pion FFs, see Figure~\ref{fig:DSS_vs_KKKS}.

\subsection{Comparison with LHCb and ALICE data}
\label{sec:ComparisonwithLHCbdata}

\begin{figure}[htb!]
\centering
\includegraphics[width=0.490\textwidth]{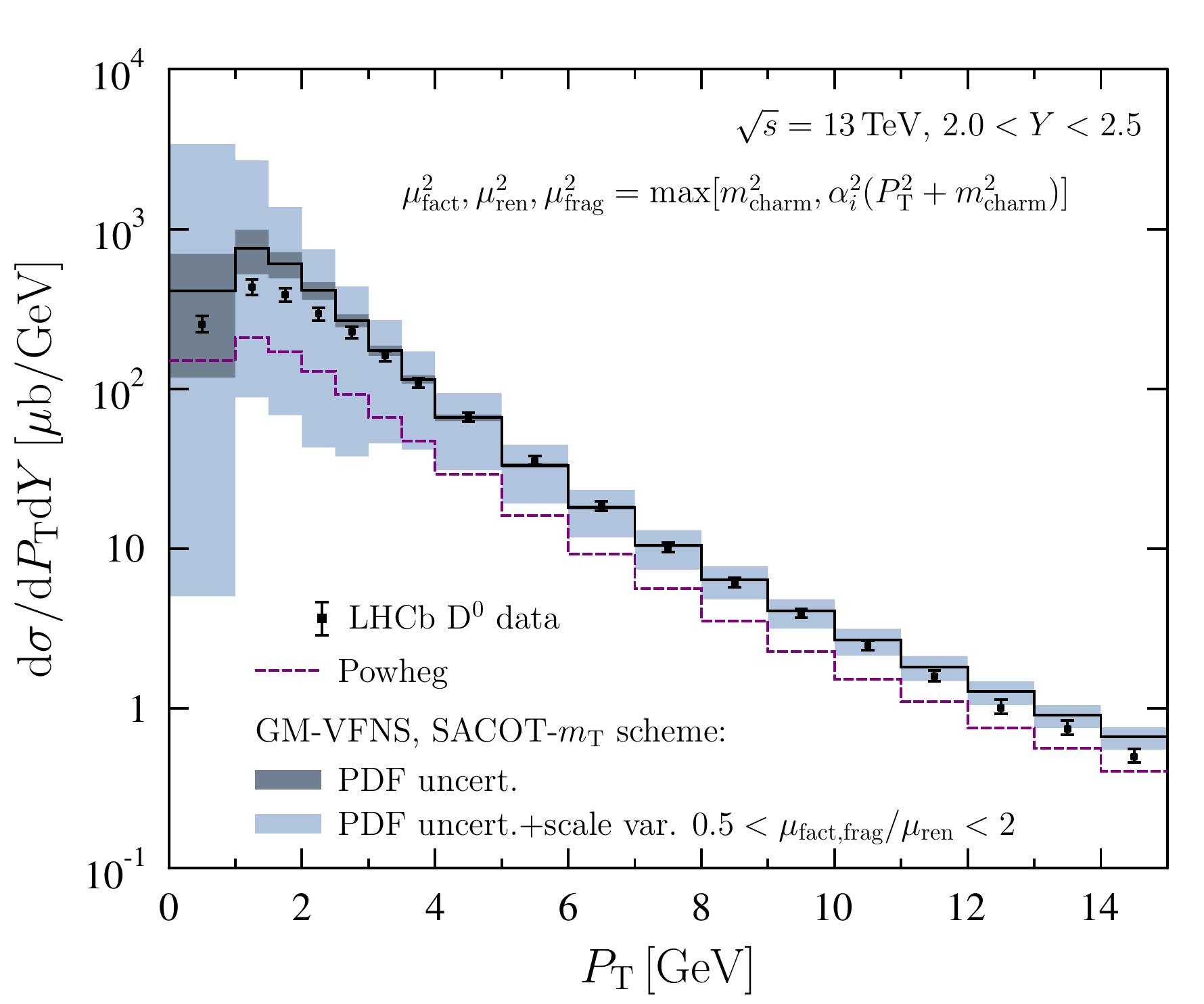}
\includegraphics[width=0.490\textwidth]{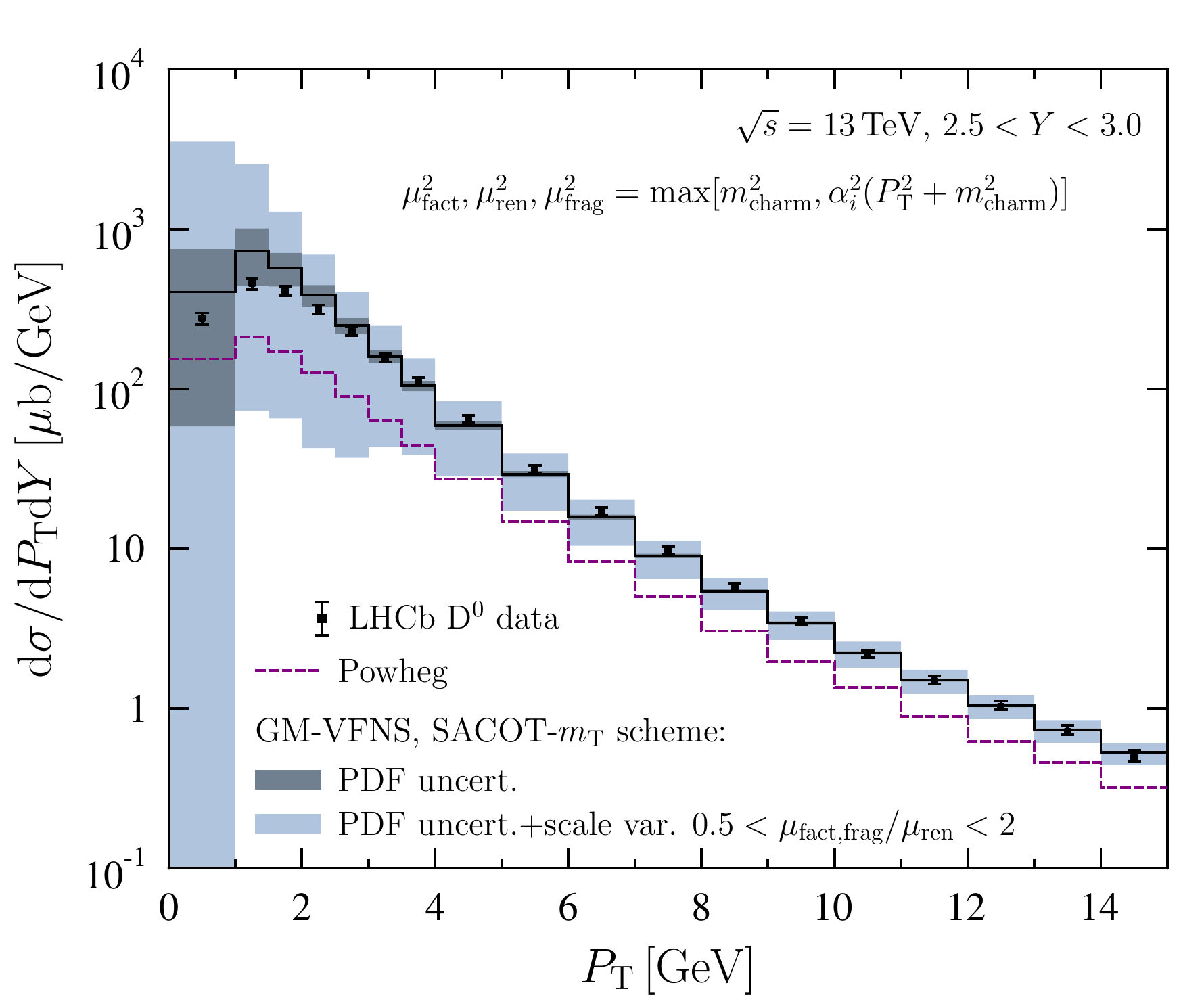}
\vspace{0.2cm}
\includegraphics[width=0.490\textwidth]{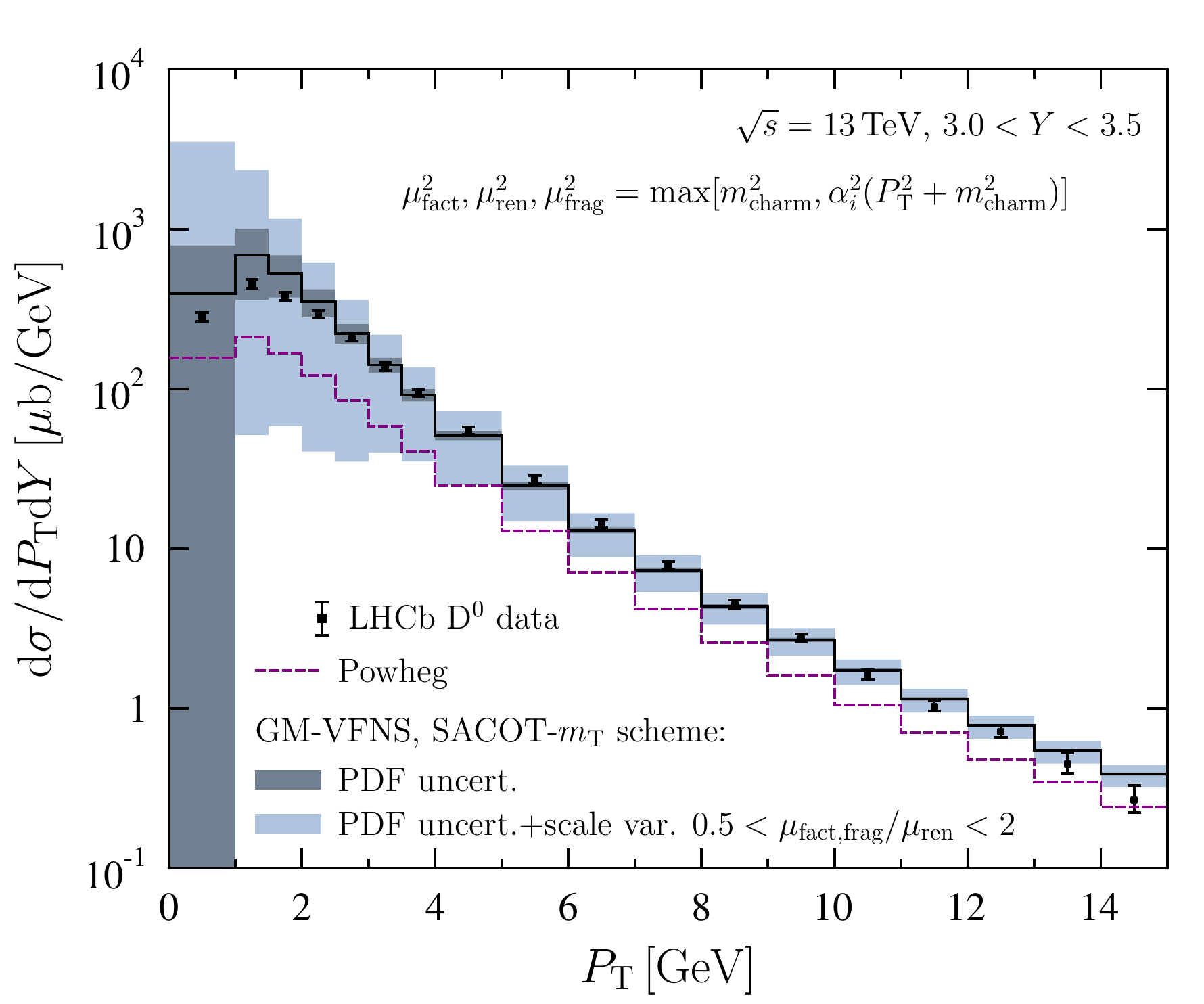}
\includegraphics[width=0.490\textwidth]{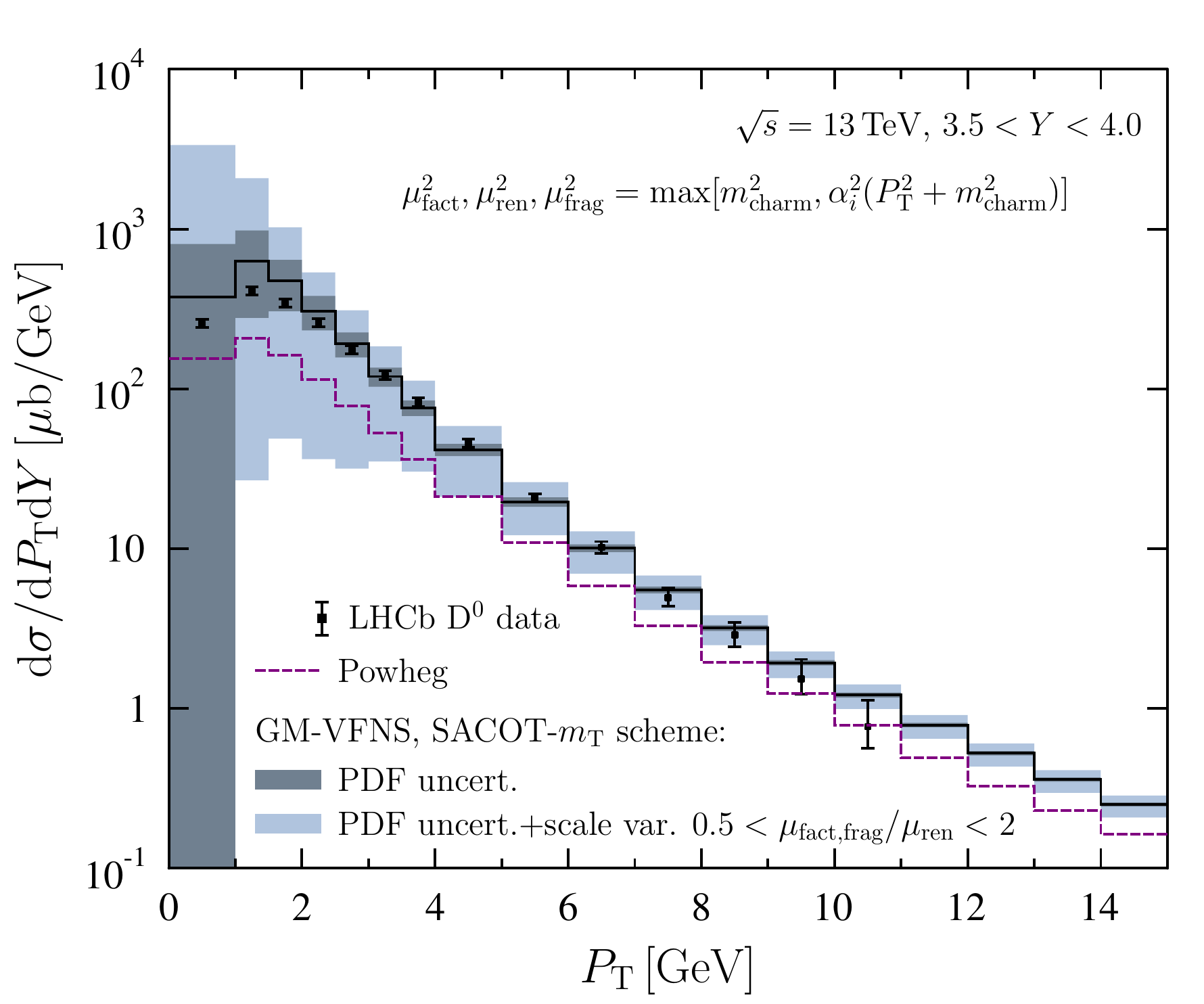}
\vspace{0.2cm}
\includegraphics[width=0.490\textwidth]{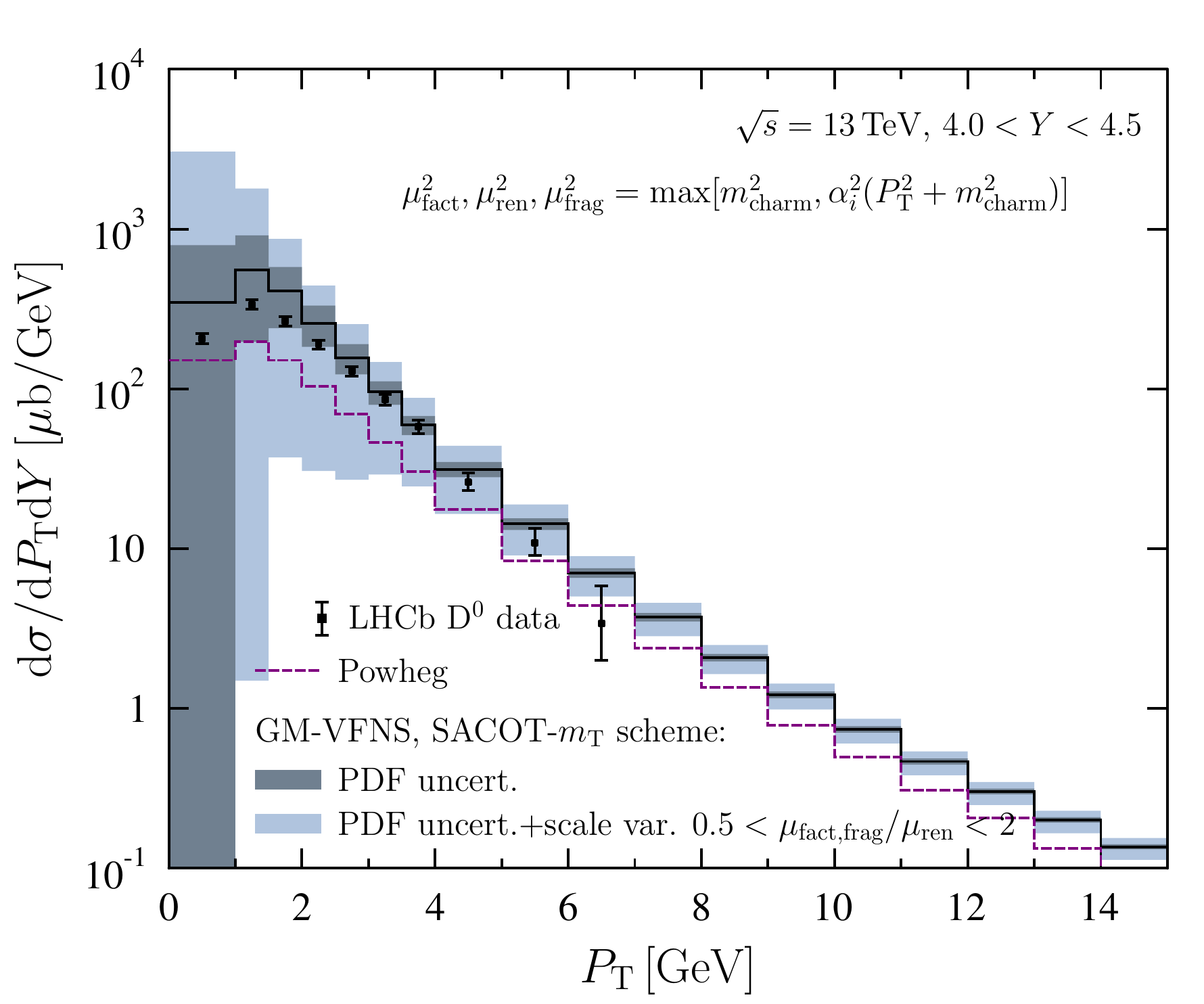}
\caption{A comparison of the LHCb $\sqrt{s}=13\, {\rm TeV}$ ${\rm D}^0+\overline{{\rm D}^0}$ data \cite{Aaij:2015bpa} with the GM-VFNS and \textsc{Powheg} calculations. The black lines indicate the central GM-VFNS results, the PDF uncertainty is shown as darker band, and the light-blue band is the PDF uncertainty and scale variation added in quadrature. The purple dashed line is the central result from \textsc{Powheg} calculation.} 
\label{fig:LHCb13}
\end{figure}
\begin{figure}[htb!]
\centering
\includegraphics[width=0.490\textwidth]{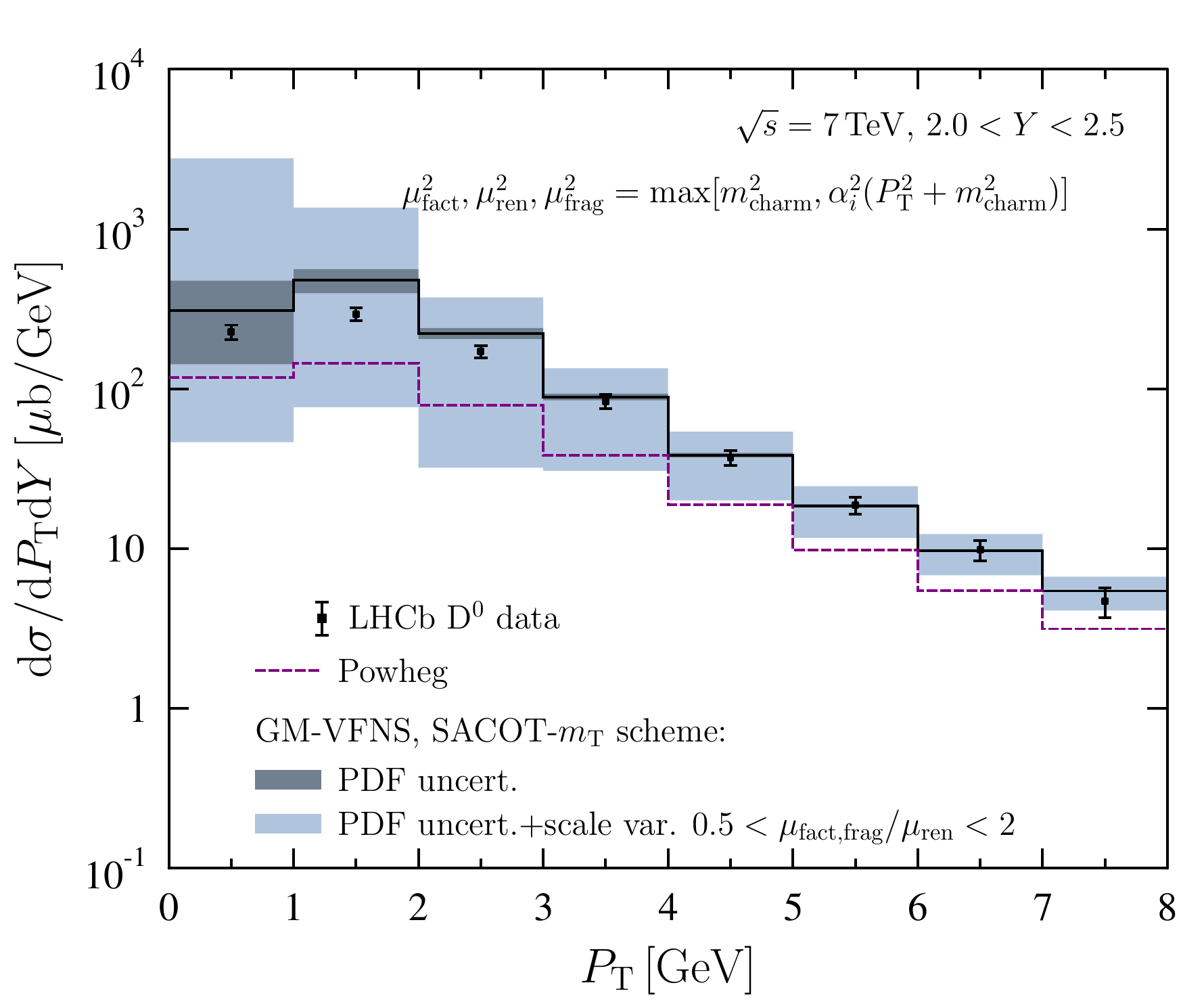}
\includegraphics[width=0.490\textwidth]{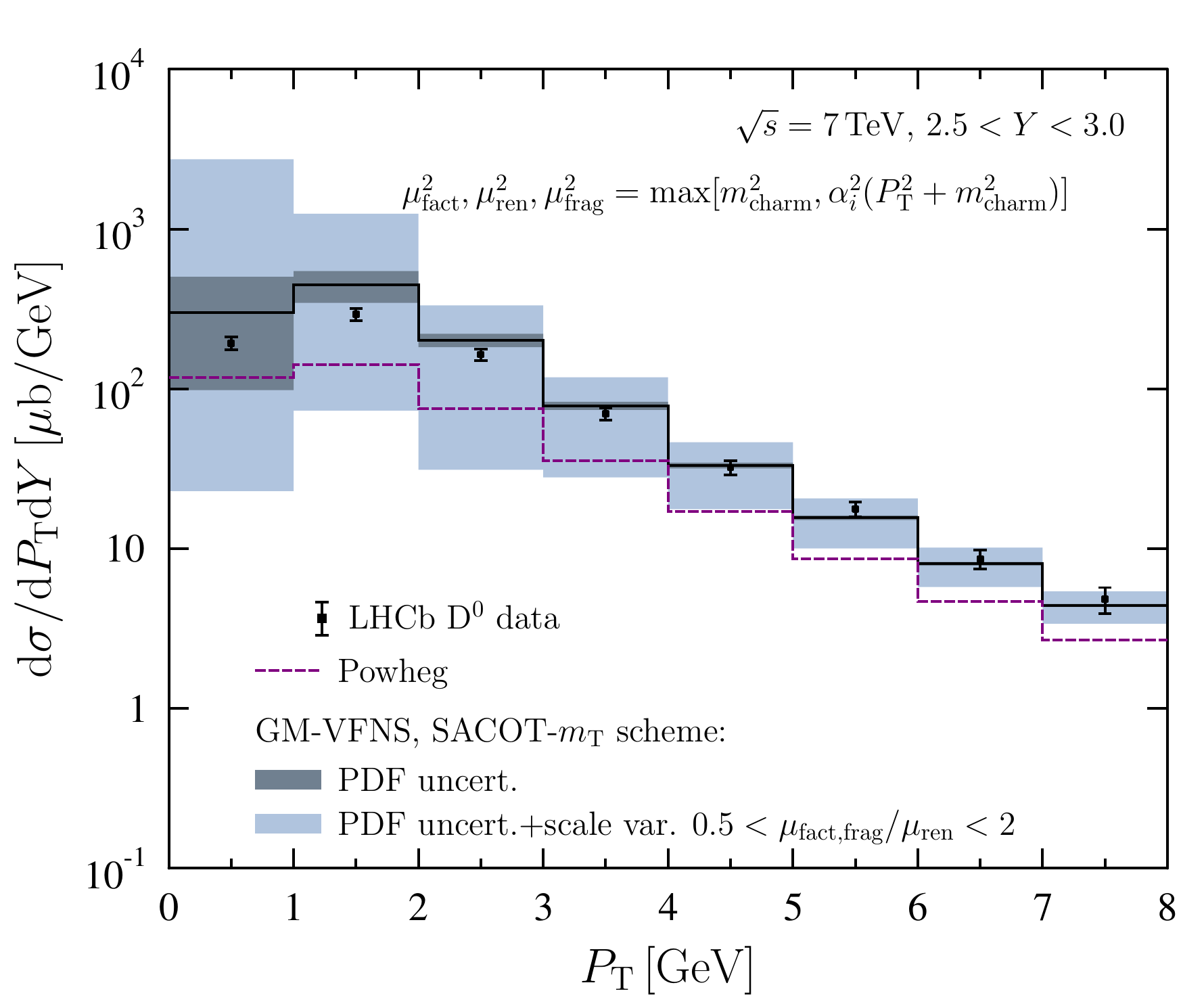}
\vspace{0.2cm}
\includegraphics[width=0.490\textwidth]{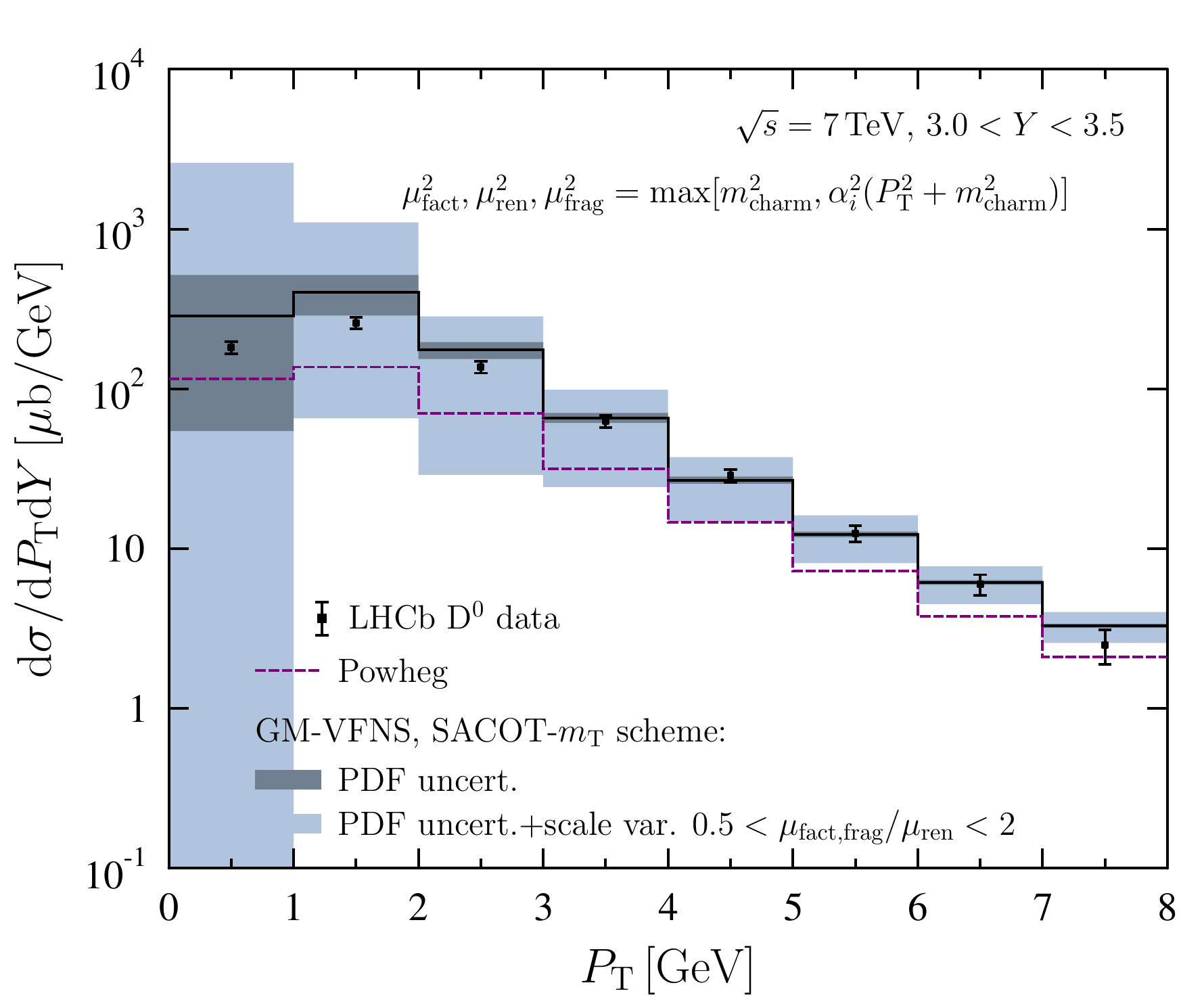}
\includegraphics[width=0.490\textwidth]{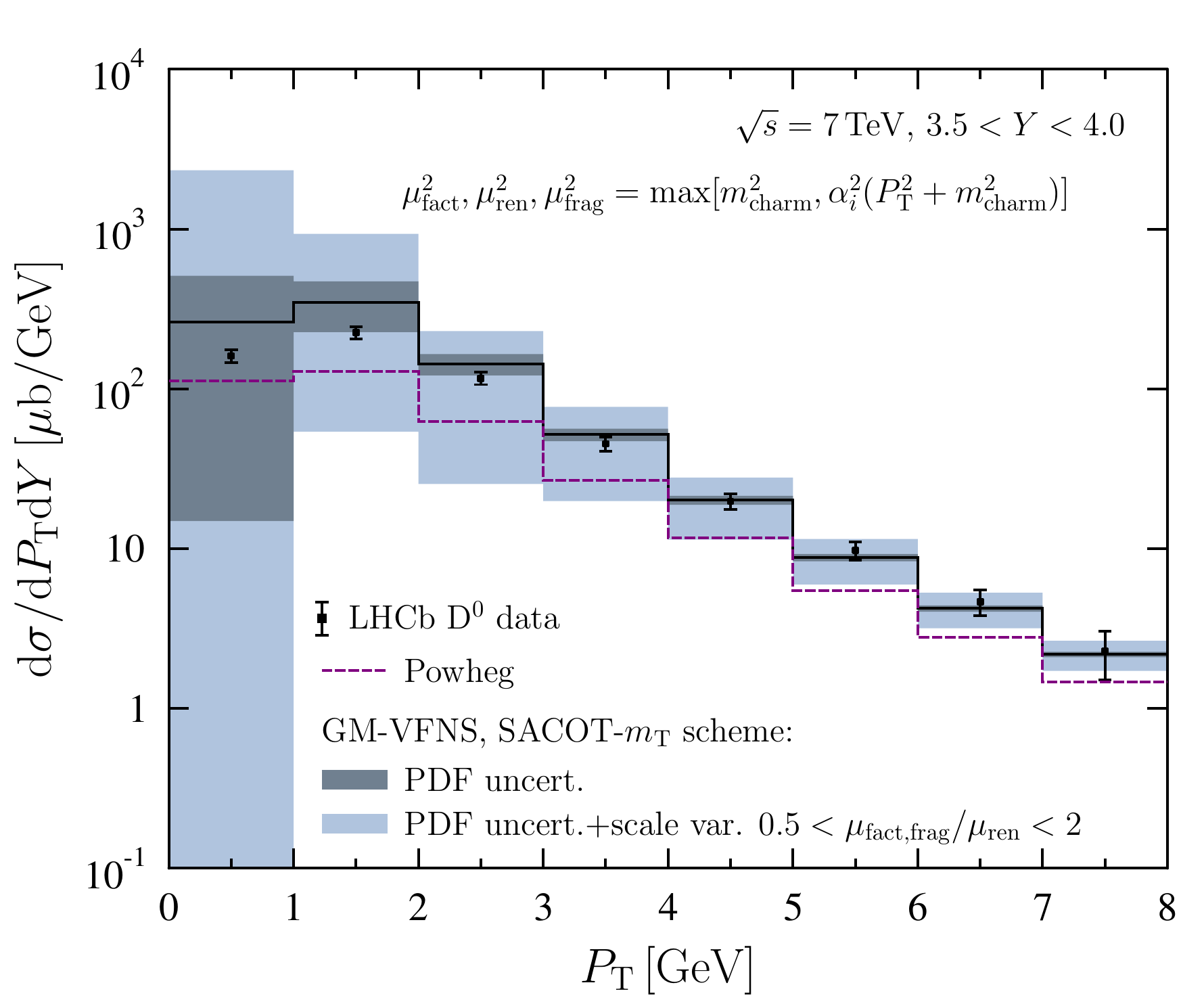}
\vspace{0.2cm}
\includegraphics[width=0.490\textwidth]{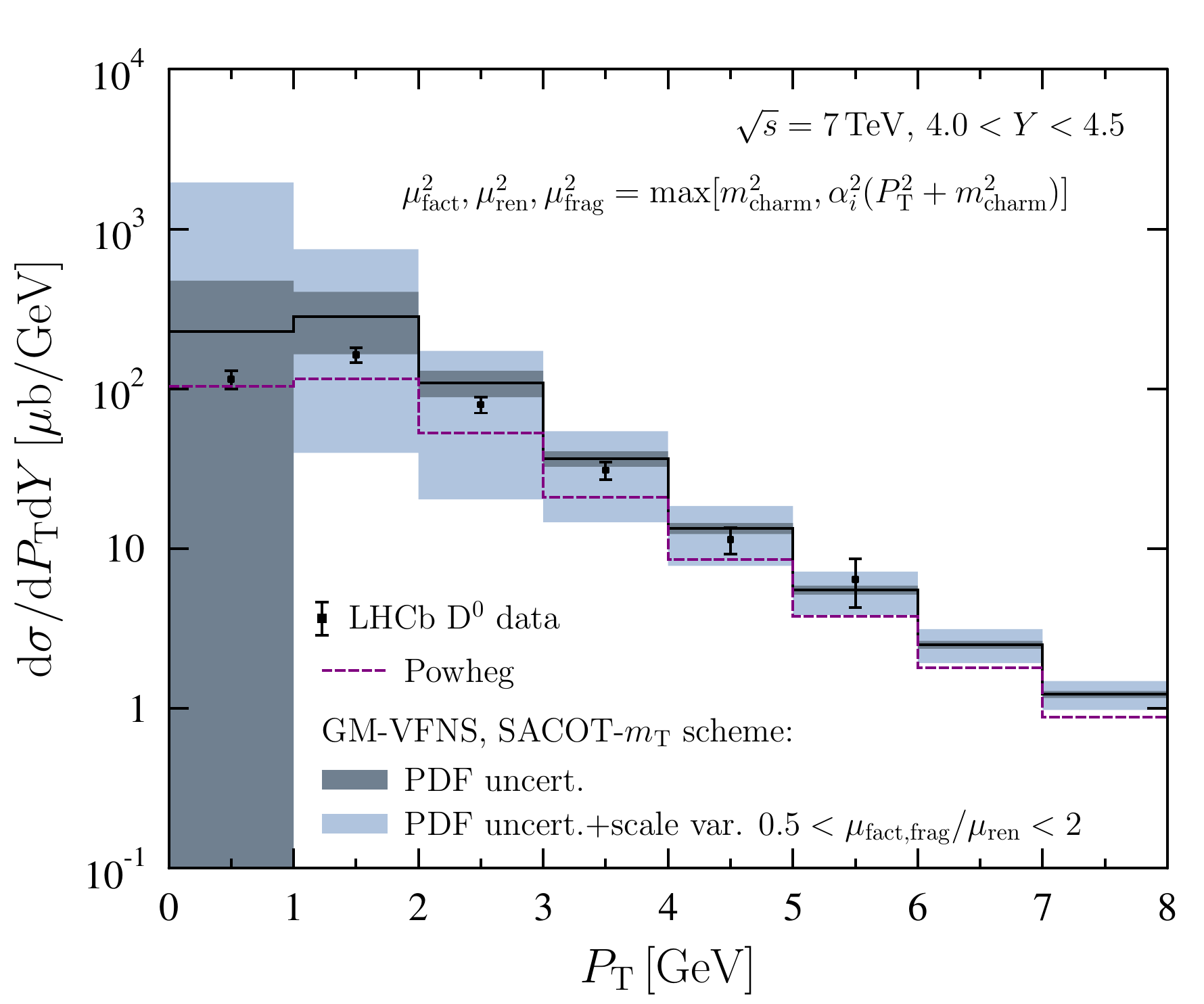}
\caption{A comparison of the LHCb $\sqrt{s}=7\, {\rm TeV}$ ${\rm D}^0+\overline{{\rm D}^0}$ data \cite{Aaij:2013mga} with the GM-VFNS and \textsc{Powheg} calculations. The black lines indicate the central GM-VFNS results, the PDF uncertainty is shown as darker band, and the light-blue band is the PDF uncertainty and scale variation added in quadrature. The purple dashed line is the central result from \textsc{Powheg} calculation.} 
\label{fig:LHCb7}
\end{figure}
\begin{figure}[htb!]
\centering
\includegraphics[width=0.490\textwidth]{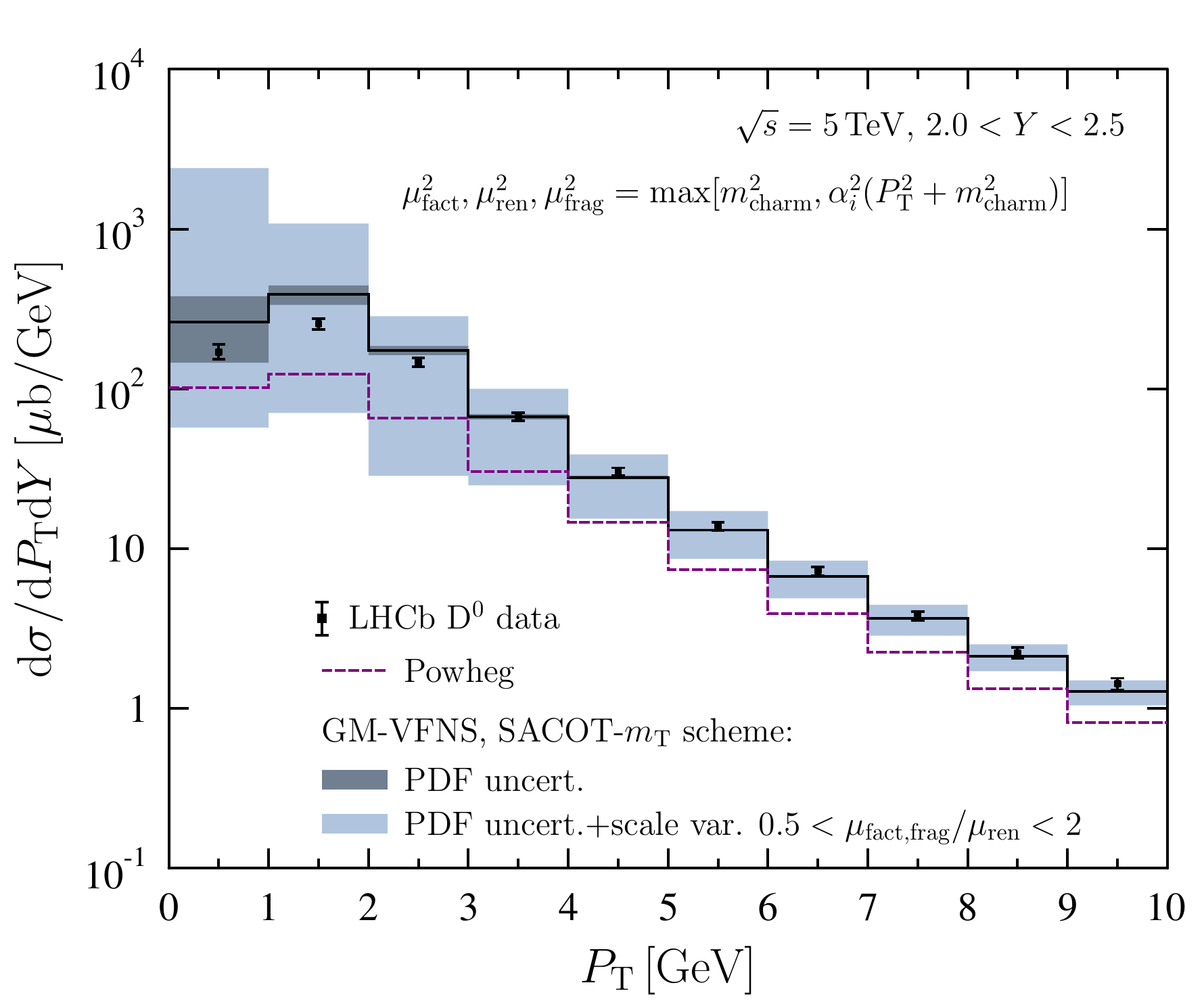}
\includegraphics[width=0.490\textwidth]{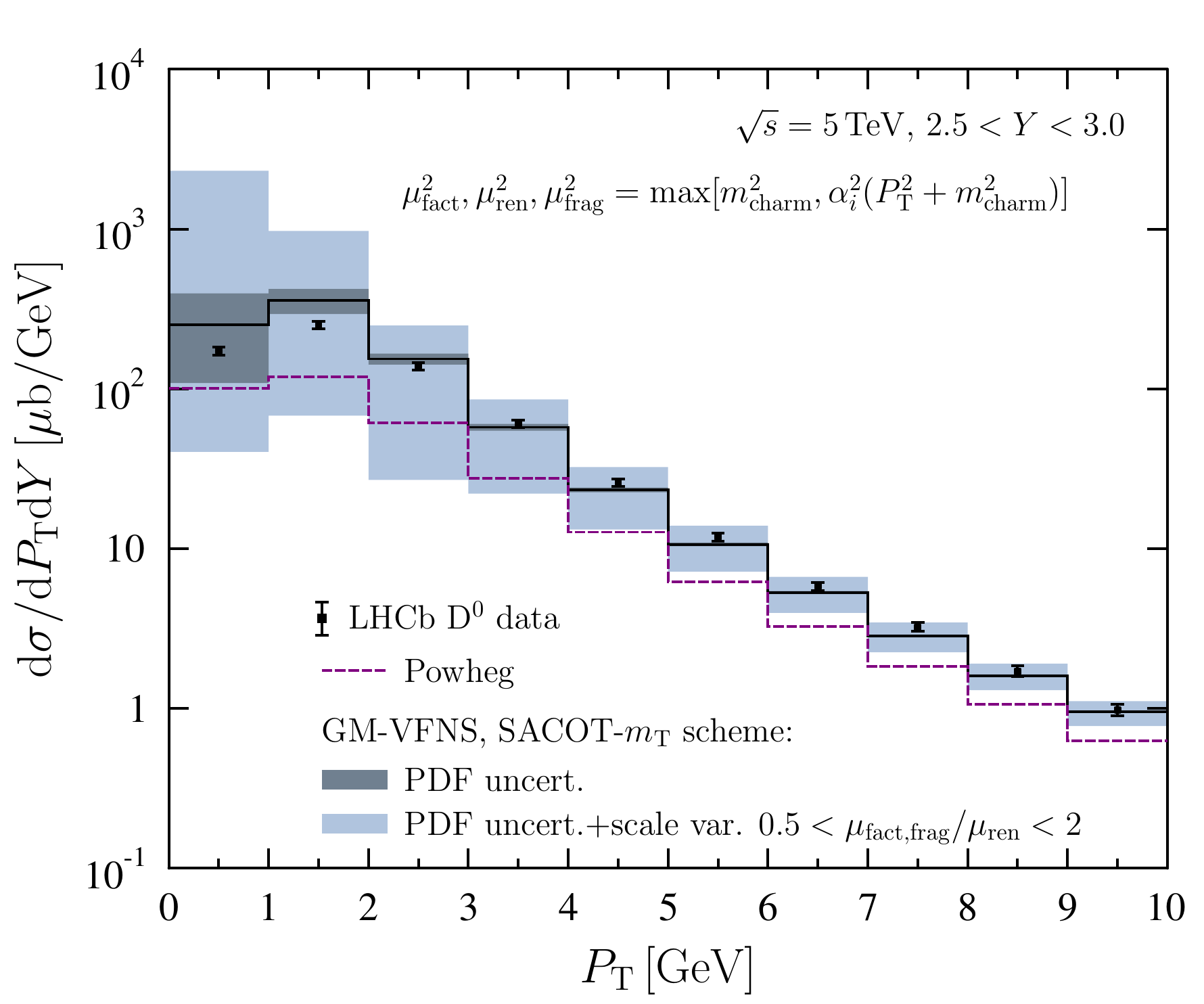}
\vspace{0.2cm}
\includegraphics[width=0.490\textwidth]{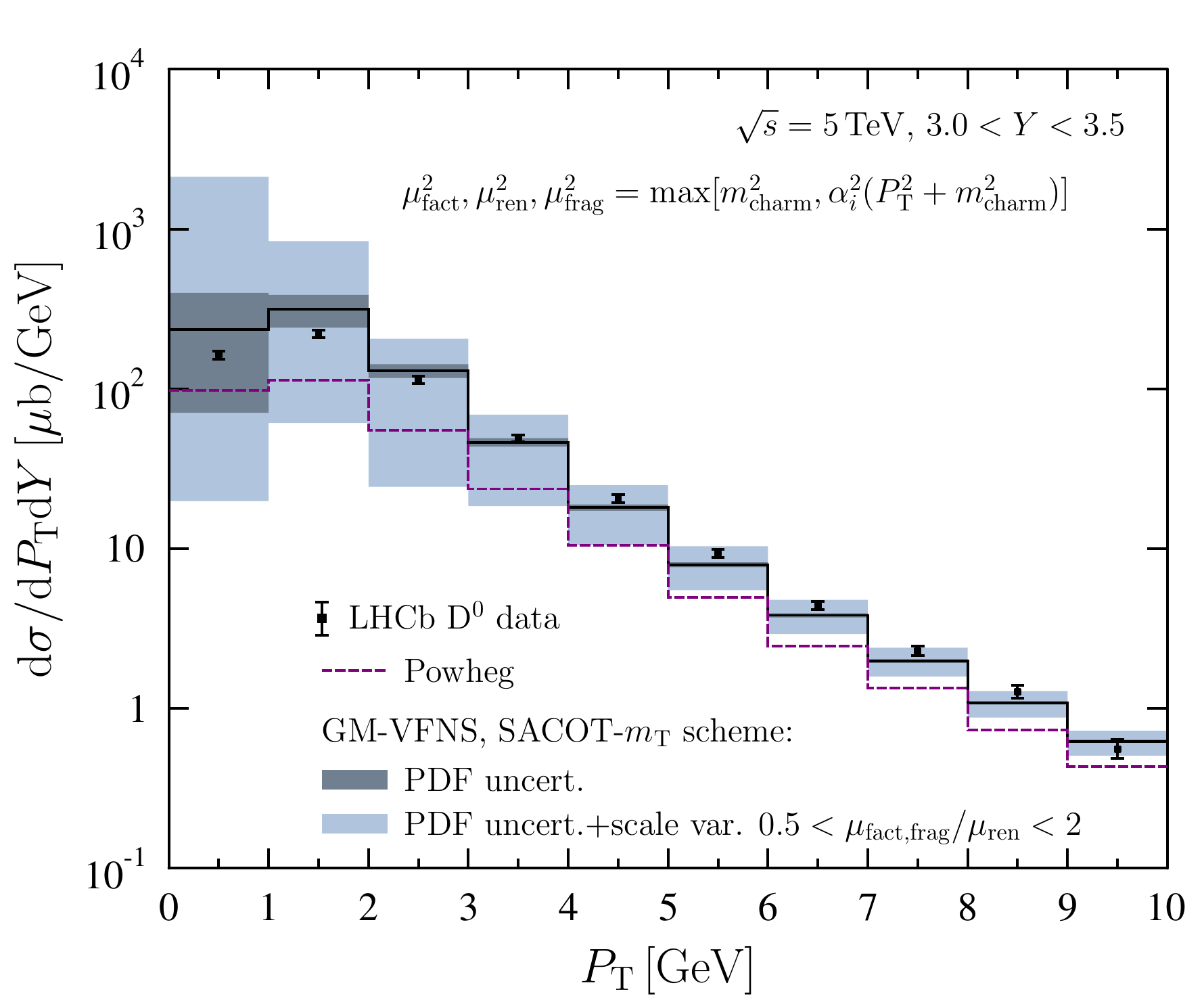}
\includegraphics[width=0.490\textwidth]{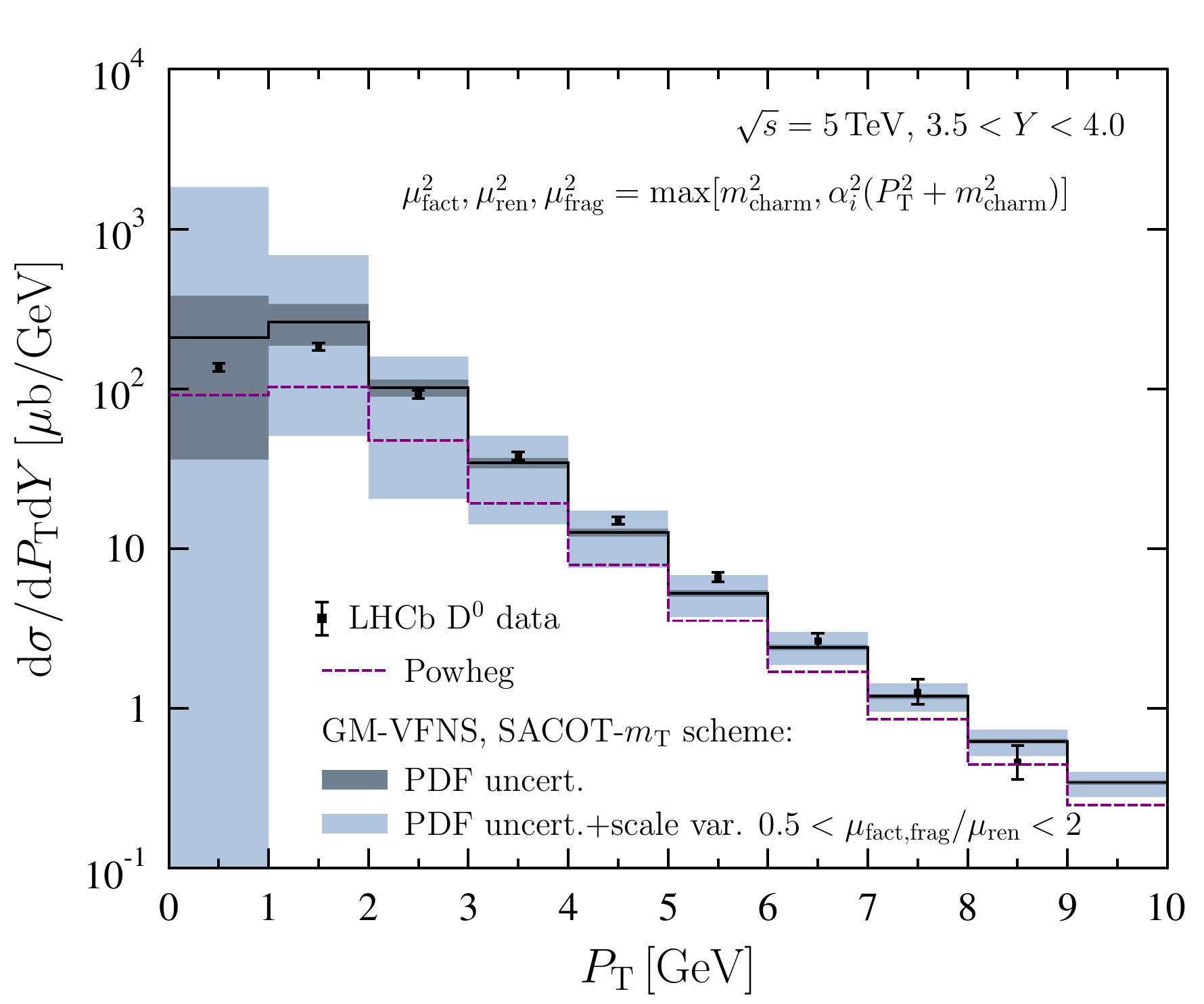}
\vspace{0.2cm}
\includegraphics[width=0.490\textwidth]{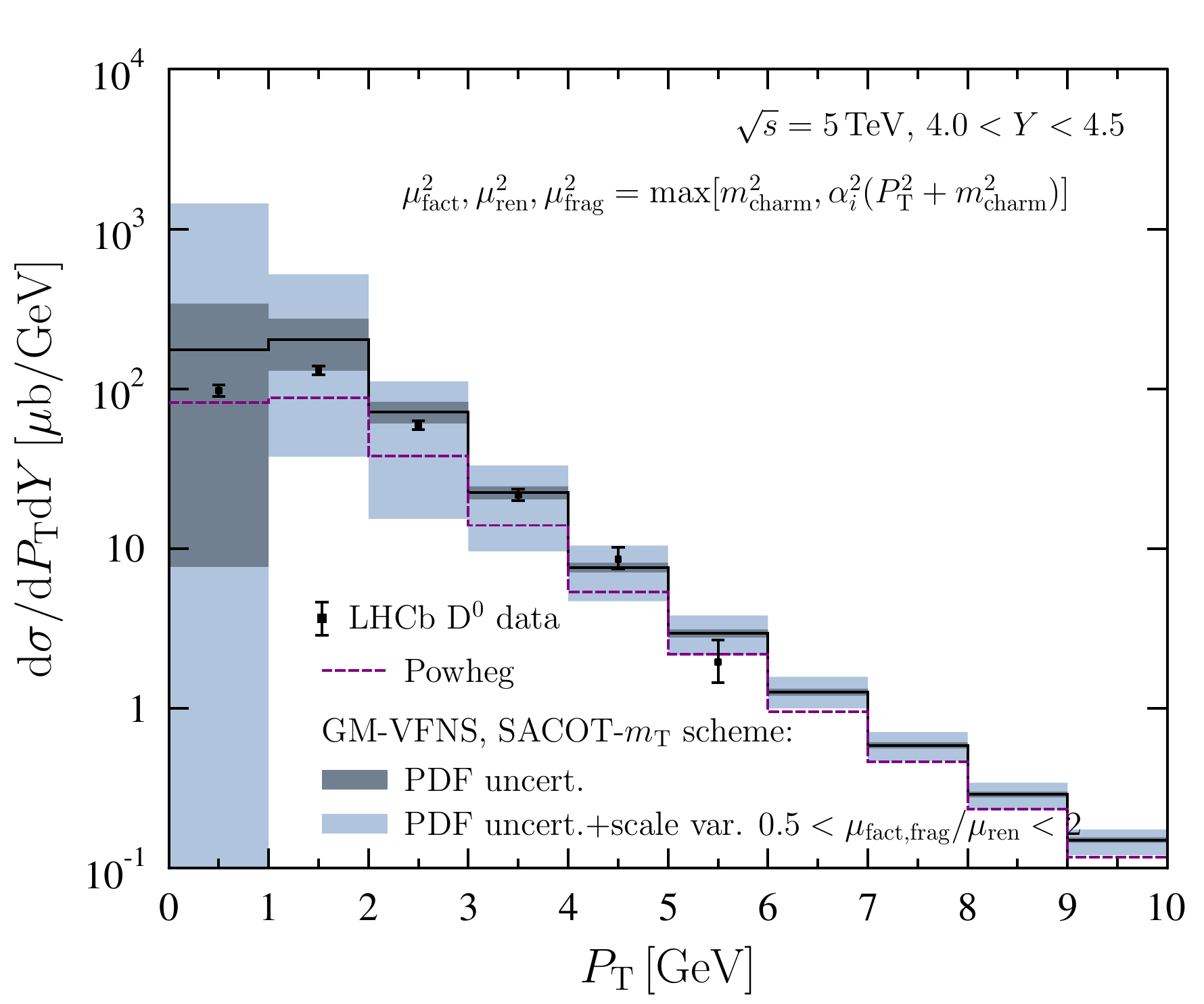}
\caption{A comparison of the LHCb $\sqrt{s}=5\, {\rm TeV}$ ${\rm D}^0+\overline{{\rm D}^0}$ data \cite{Aaij:2016jht} with the GM-VFNS and \textsc{Powheg} calculations. The black lines indicate the central GM-VFNS results, the PDF uncertainty is shown as darker band, and the light-blue band is the PDF uncertainty and scale variation added in quadrature. The purple dashed line is the central result from \textsc{Powheg} calculation.} 
\label{fig:LHCb5}
\end{figure}
\begin{figure}[htb!]
\centering
\includegraphics[width=0.490\textwidth]{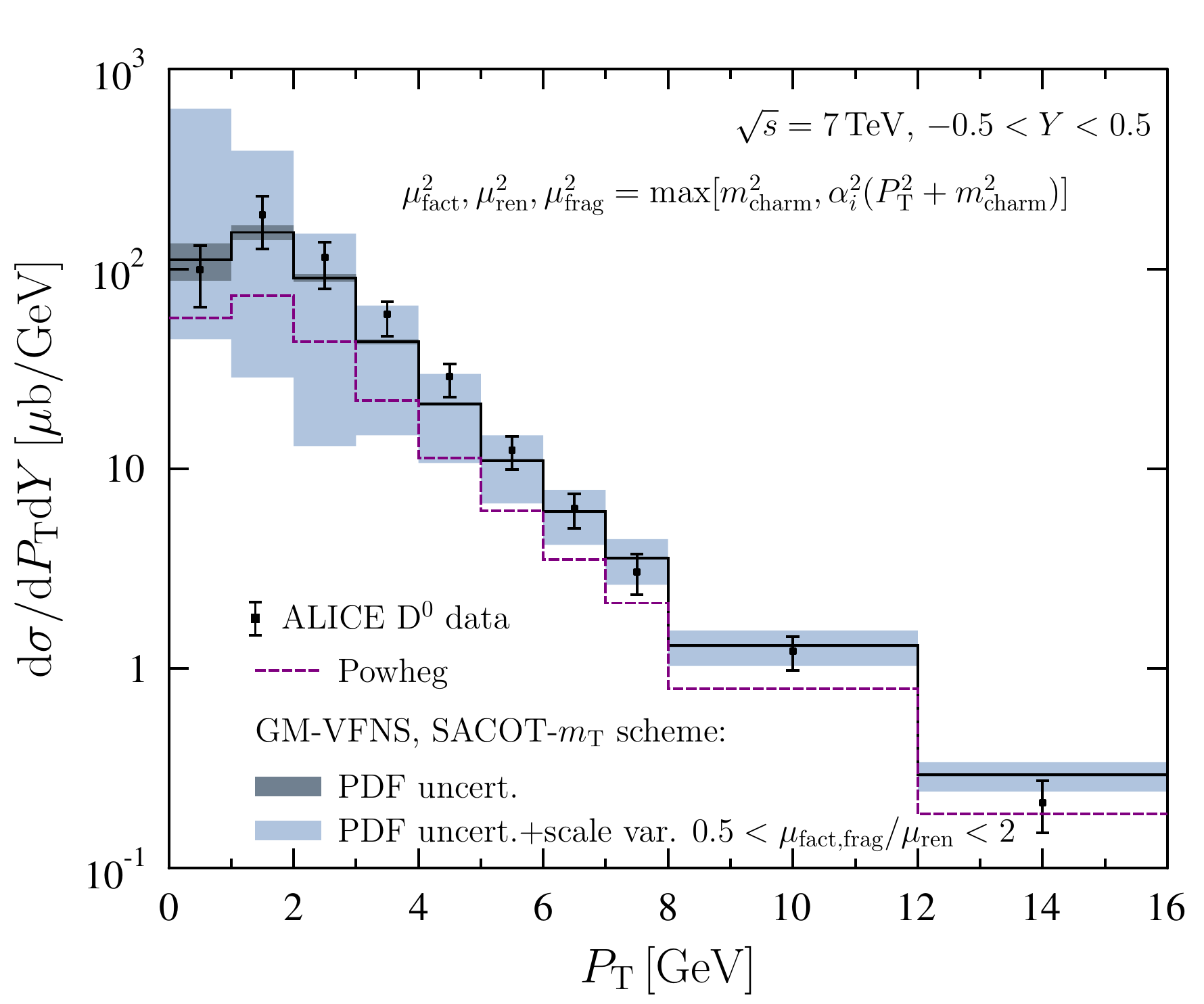}
\caption{A comparison of the ALICE $\sqrt{s}=7\, {\rm TeV}$ $({\rm D}^0 + \overline{{\rm D}^0})/2$ data \cite{Adam:2016ich} with the GM-VFNS and \textsc{Powheg} calculations. The black lines indicate the central GM-VFNS results, the PDF uncertainty is shown as darker band, and the light-blue band is the PDF uncertainty and scale variation added in quadrature. The purple dashed line is the central result from \textsc{Powheg} calculation. The experimental 3.5\% luminosity and 1.3\% global uncertainty on $D^0 \rightarrow K^-\pi^+$ branching ratio are not shown.
} 
\label{fig:ALICE7}
\end{figure}

\begin{figure}[htb!]
\centering
\includegraphics[width=0.490\textwidth]{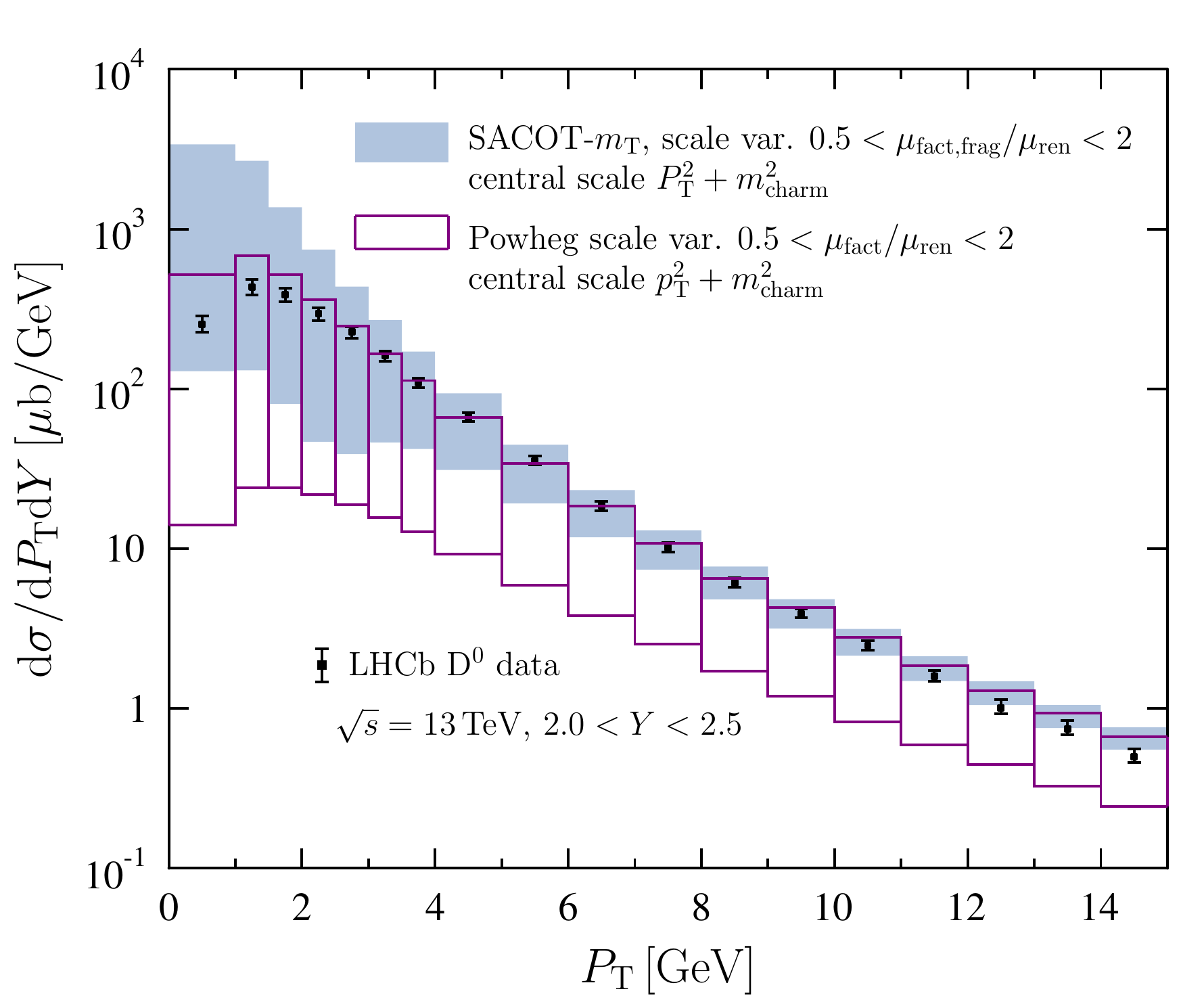}
\includegraphics[width=0.490\textwidth]{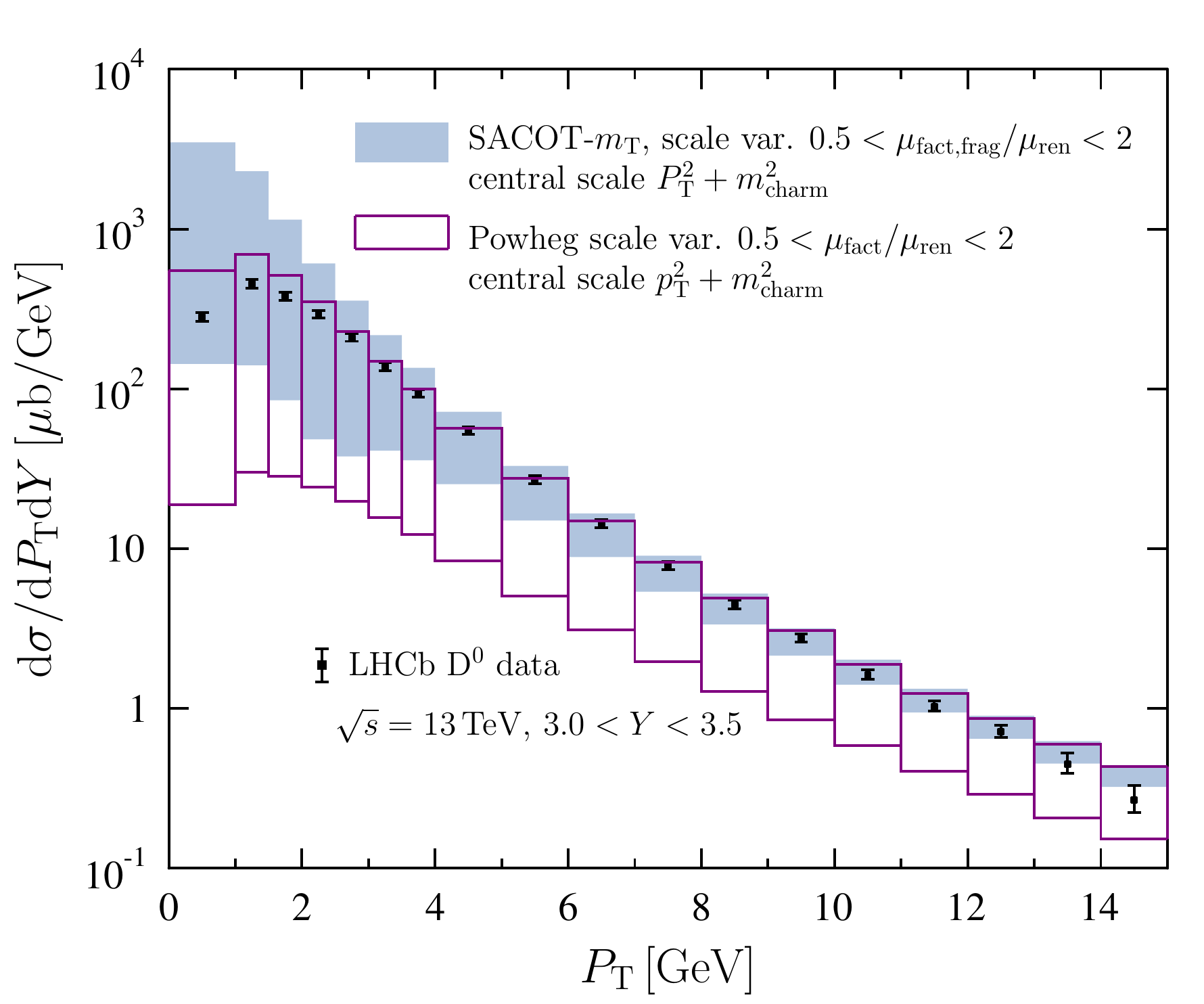}
\caption{A comparison of scale uncertainties between GM-VFNS (light-blue band) and \textsc{Powheg} (purple boxes) calculations for ${\rm D}^0+\overline{{\rm D}^0}$ production at $\sqrt{s}=13\, {\rm TeV}$ for rapidity bins $Y\in [2.0,2.5]$ (left) and $Y\in [3.0,3.5]$ (right). Also the data from LHCb \cite{Aaij:2015bpa} are shown as in Figure~\ref{fig:LHCb13}.}
\label{fig:GM-VFNS_vs_PP}
\end{figure}

In this section we will present comparisons of our calculation with the experimental LHCb and ALICE D$^0$ data taken in p-p collisions at $\sqrt{s}=5\, {\rm GeV}$ \cite{Aaij:2016jht}, $\sqrt{s}=7\, {\rm GeV}$ \cite{Aaij:2013mga,Adam:2016ich} and $\sqrt{s}=13\, {\rm GeV}$ \cite{Aaij:2015bpa}. We will consider two types of theoretical uncertainties, namely those related to the choices of the QCD scales and those related to PDF errors. In principle, there are also other sources of theory uncertainty in the input variables, like the value picked for the charm-quark mass or the value of $\alpha_s$ at the Z-boson pole. However, to consistently vary these quantities they should be accompanied by PDFs and FFs extracted with the same variation. The are no error sets (unlike for most modern PDFs) available for the D$^0$ meson fragmentation functions, so the uncertainties in FFs are not considered either. The definitions of fragmentation variable and heavy-quark scheme are taken here as inherent to the presented calculation.

The PDF uncertainties are evaluated in the standard NNPDF way by computing the variance
\begin{equation}
 \delta \langle \mathrm{d}\sigma \rangle = \sqrt{\frac{1}{N_{\rm rep}} \sum_i \left(\mathrm{d}\sigma_i-\langle \mathrm{d}\sigma \rangle \right)^2},
\end{equation}
where $d\sigma_i$ is the cross section computed with the $i$th member out of the collection of $N_{\rm rep}$ PDF replicas, and where
\begin{equation}
 \langle \mathrm{d}\sigma \rangle = \frac{1}{N_{\rm rep}} \sum_i \mathrm{d}\sigma_i  \,,
\end{equation}
is the central prediction. The stability of the results against scale variations is quantified as e.g. in Ref.~\cite{dEnterria:2013sgr} by varying the three scales independently as
\begin{equation}
\mu_{\rm ren}, \mu_{\rm fact}, \mu_{\rm frag} = \max\left(m_{\rm charm}, \alpha_i \sqrt{P_{\rm T}^2 + m_{\rm charm}^2} \right), \quad 0.5 < \frac{\mu_{\rm fact}, \mu_{\rm frag}}{\mu_{\rm ren}}  < 2 \,,
\end{equation}
where the parameters $\alpha_{i=\rm fact,ren, frag}$ vary between $0.5$ and $2$. The total scale uncertainty is taken as the maximum and minimum of the 16 cross section found in this way. With the above choice, the scales remain always above the charm mass-threshold and the potentially large contributions from $\log(\mu_{\rm ren}/\mu_{\rm fact})$ and $\log(\mu_{\rm ren}/\mu_{\rm frag})$ terms are suppressed by limiting the maximal difference of the respective scales to a factor of two.
The results for absolute cross sections are presented in Figure~\ref{fig:LHCb13} in the case of $\sqrt{s}=13 \, {\rm TeV}$ LHCb p-p data, Figure~\ref{fig:LHCb7} in the case of $\sqrt{s}=7 \, {\rm TeV}$ LHCb p-p data, Figure~\ref{fig:LHCb5} in the case of $\sqrt{s}=5 \, {\rm TeV}$ LHCb p-p data, and Figure~\ref{fig:ALICE7} in the case of $\sqrt{s}=7 \, {\rm TeV}$ ALICE p-p data. In addition to our GM-VFNS results, we also show the central prediction from the \textsc{Powheg}+\textsc{Pythia} framework in all panels, and separately compare the scale uncertainties of these two different approaches in two rapidity bins at $\sqrt{s}=13 \, {\rm TeV}$ in Figure~\ref{fig:GM-VFNS_vs_PP}. 

In all cases, the data are reproduced very well by the GM-VFNS calculations within the considered theory uncertainties, whereas the central values of the \textsc{Powheg}+\textsc{Pythia} calculations systematically fall short of the data. Essentially the same hierarchy has been observed e.g. in LHCb/ALICE papers \cite{Aaij:2016jht,Aaij:2013mga,Adam:2016ich,Aaij:2015bpa} and elsewhere \cite{Klasen:2014dba}, though the GM-VFNS calculations do not extend to zero $P_{\rm T}$ in these references. 
\begin{figure}[tbhp]
\centering
\includegraphics[width=0.50\textwidth]{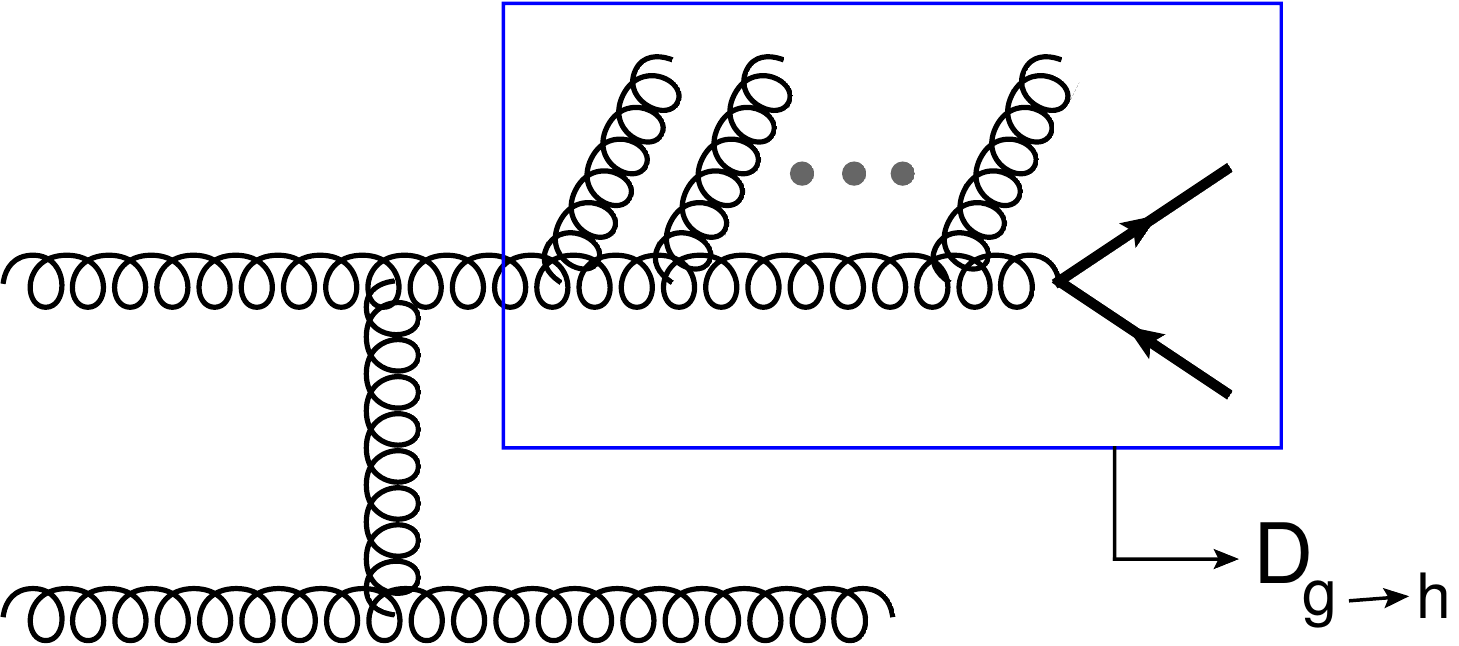}
\caption{A contribution to heavy-quark production resummed by DGLAP equations in GM-VFNS calculation. The diagram has been drawn with JaxoDraw 2.1 \cite{Binosi:2003yf}.} 
\label{fig:fig2}
\end{figure}
We believe that most, or at least a significant part, of the systematic difference between GM-VFNS and \textsc{Powheg}+\textsc{Pythia} setups can be explained by contributions from gluon fragmentation that are resummed in GM-VFNS but that are not accounted for in the \textsc{Powheg}+\textsc{Pythia} calculation. To better illustrate the point we show, in Figure~\ref{fig:fig2}, a diagram where the upper gluon line radiates one or more gluons before splitting into a $Q\overline{Q}$ pair. Contributions of this type are implicitly included in the GM-VFNS calculation, resummed into the gluon FF $D_{g \rightarrow h_3}$ in collinear configuration. They correspond to roughly 50\% of the $D^0$-meson cross sections in $P_{\mathrm{T}}$ and $Y$ region investigated here (ALICE and LHCb acceptance). The \textsc{Powheg}+\textsc{Pythia} framework, however, does not allow for these contributions as the starting point are events in which the $Q\overline{Q}$ pair has been produced in the first place. In other words, the possibility that the $Q\overline{Q}$ pair is produced only later on by the \textsc{Pythia} parton shower is possible only for the hard processes where one $Q\overline{Q}$ pair has already been produced in matrix-element level. On the other hand, the standard \textsc{Pythia} simulation with all hard-QCD processes subsumed does include also contributions like those in Figure~\ref{fig:fig2} and, as was shown in the lower panel of Figure~\ref{fig:xdists}, enabling this possibility changes the $x_2$ distributions quite significantly. In the FONLL approach \cite{Cacciari:1998it}, the contributions like those in Figure~\ref{fig:fig2} are resummed by partonic FFs, but the contribution of the resummed part is shrouded by a multiplicative factor $G(m,p_{\rm T})=p_{\rm T}^2/(p_{\rm T}^2 + 25m^2)$ engineered so as to suppress these contributions at low $p_{\rm T}$.

\begin{figure}[htb!]
\centering
\includegraphics[width=0.66\textwidth]{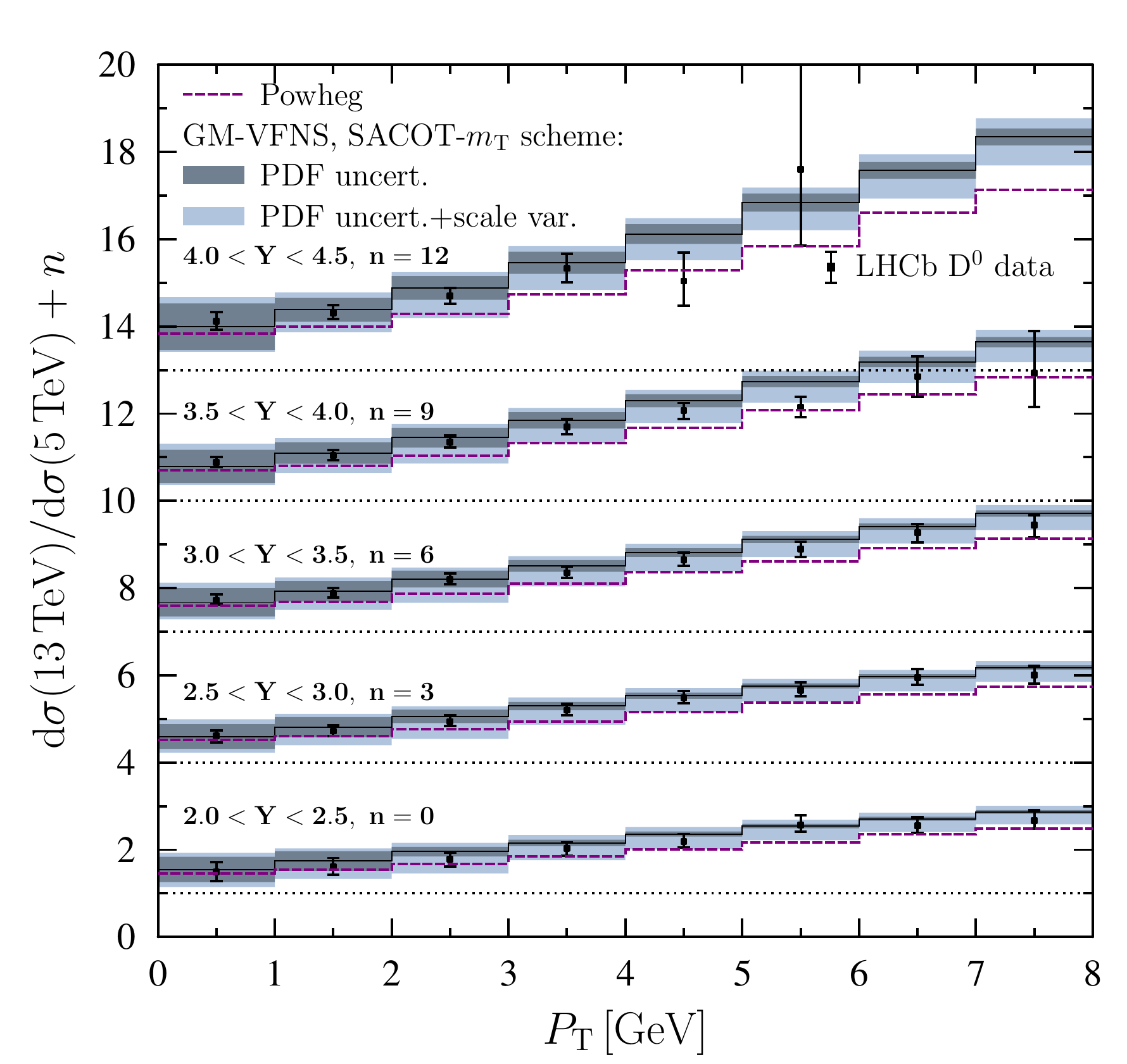}
\includegraphics[width=0.66\textwidth]{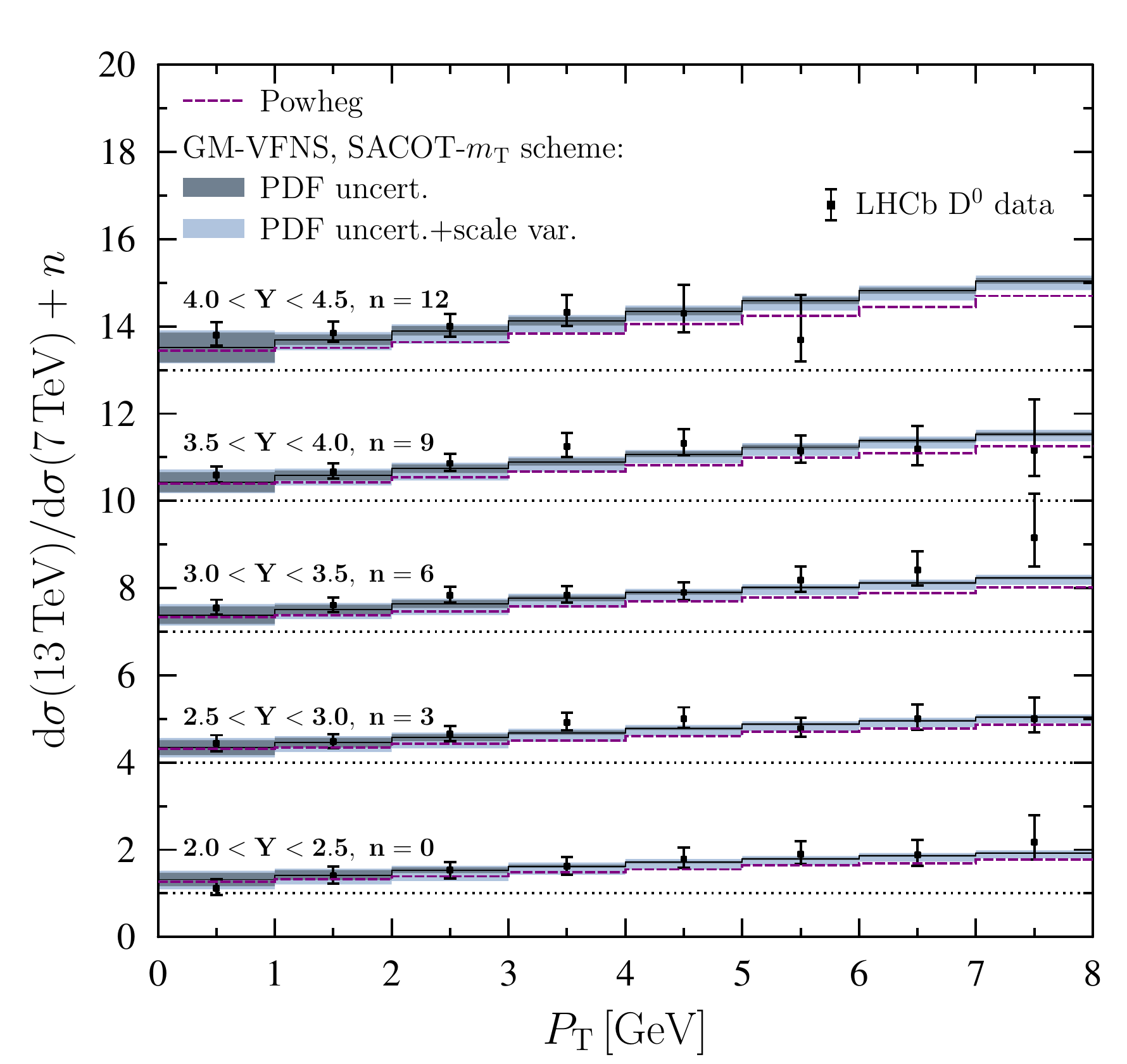}
\caption{A comparison of the LHCb ${\rm D}^0+\overline{{\rm D}^0}$ data \cite{Aaij:2015bpa,Aaij:2016jht} for cross-section ratios with the GM-VFNS and \textsc{Powheg} calculations. The black lines indicate the central GM-VFNS results, the PDF uncertainties are shown as darker bands, and the light-blue band is the PDF uncertainty and scale variation added in quadrature. The purple dashed line is the central result from \textsc{Powheg} calculation.} 
\label{fig:LHCbratios}
\end{figure}

As expected, the scale variation in GM-VFNS calculations, best visible in Figure~\ref{fig:GM-VFNS_vs_PP}, is maximal in the low-$P_{\rm T}$ region and diminishes toward larger values. At the LHCb acceptance, the scale uncertainty is clearly larger than the variation we found using massless/massive fragmentation variable, see back Figure~\ref{fig:systeff}. At low $P_{\rm T}$, the upper limit for the scale variation is set by the calculation with $\alpha_{\rm fact,frag}=2$ and $\alpha_{\rm ren}=1$, and the lower border traces the configuration $\alpha_{\rm fact,frag}=0.5$ and $\alpha_{\rm ren}=1$. 
Neither of these ``extreme'' cases reproduces the shape of the data particularly well at low $P_{\rm T}$: either the spectra rise too steeply ($\alpha_{\rm fact,frag}=2$)\footnote{The turndowns towards zero do take place, but so close to $P_{\rm T}=0$ that they are not visible within the used $P_{\rm T}$ binning.}, or they are too flat ($\alpha_{\rm fact,frag}=0.5$). The change of slope of the lower border between $P_{\rm T} \sim 3 \, {\rm GeV}$ and $P_{\rm T} \sim 5 \, {\rm GeV}$ is a consequence of the heavy-quark PDFs and light-parton FFs becoming 'active' as the scale goes above the mass threshold. That is, the contributions shown in the right-hand panel of Figure~\ref{fig:check} become suddenly dominant. Below $P_{\rm T} \sim 3 \, {\rm GeV}$ the set condition $\min(\mu_i)>m_{\rm charm}$ halts the growth of scale uncertainty downwards. The ``natural'' choice $\alpha_i=1$ matches much better with the shape of data. The central prediction goes somewhat above the LHCb data at low $P_{\rm T}$, which could be improved by using a bit lower scale. Indeed, we have checked that setting the scales as $\sqrt{(\alpha' P_{\rm T})^2 + m_{\rm charm}^2} \,,$ with the parameter $\alpha'<1$ \cite{Kniehl:2011bk}, would improve the description with the central prediction. However, here our intention is not to fine tune the predictions but rather to present the calculation as it is ``out of the box'' with default settings. In the \textsc{Powheg}+\textsc{Pythia} case the scales $\mu_{\rm ren}$ and $\mu_{\rm fact}$ are varied independently by a factor of two wrt. central scale within $0.5 < \mu_{\rm fact}/\mu_{\rm ren} < 2$ in the generation of \textsc{Powheg} $\mathrm{c}\overline{\mathrm{c}}$ events. As can be seen in Figure~\ref{fig:GM-VFNS_vs_PP}, the \textsc{Powheg}+\textsc{Pythia} approach is clearly more sensitive to the scale variations than GM-VFNS, especially at $P_{\mathrm{T}}>4\,{\rm GeV}$, and the LHCb data still remain within these uncertainties, though at the very upper part of the band. The scale-uncertainty estimates of \textsc{Powheg}+\textsc{Pythia} method we find here are well in line with the error bands shown in the original data papers of LHCb and ALICE, and e.g. in Ref.~\cite{Gauld:2015yia}.

\begin{figure}[htb!]
\centering
\includegraphics[width=0.66\textwidth]{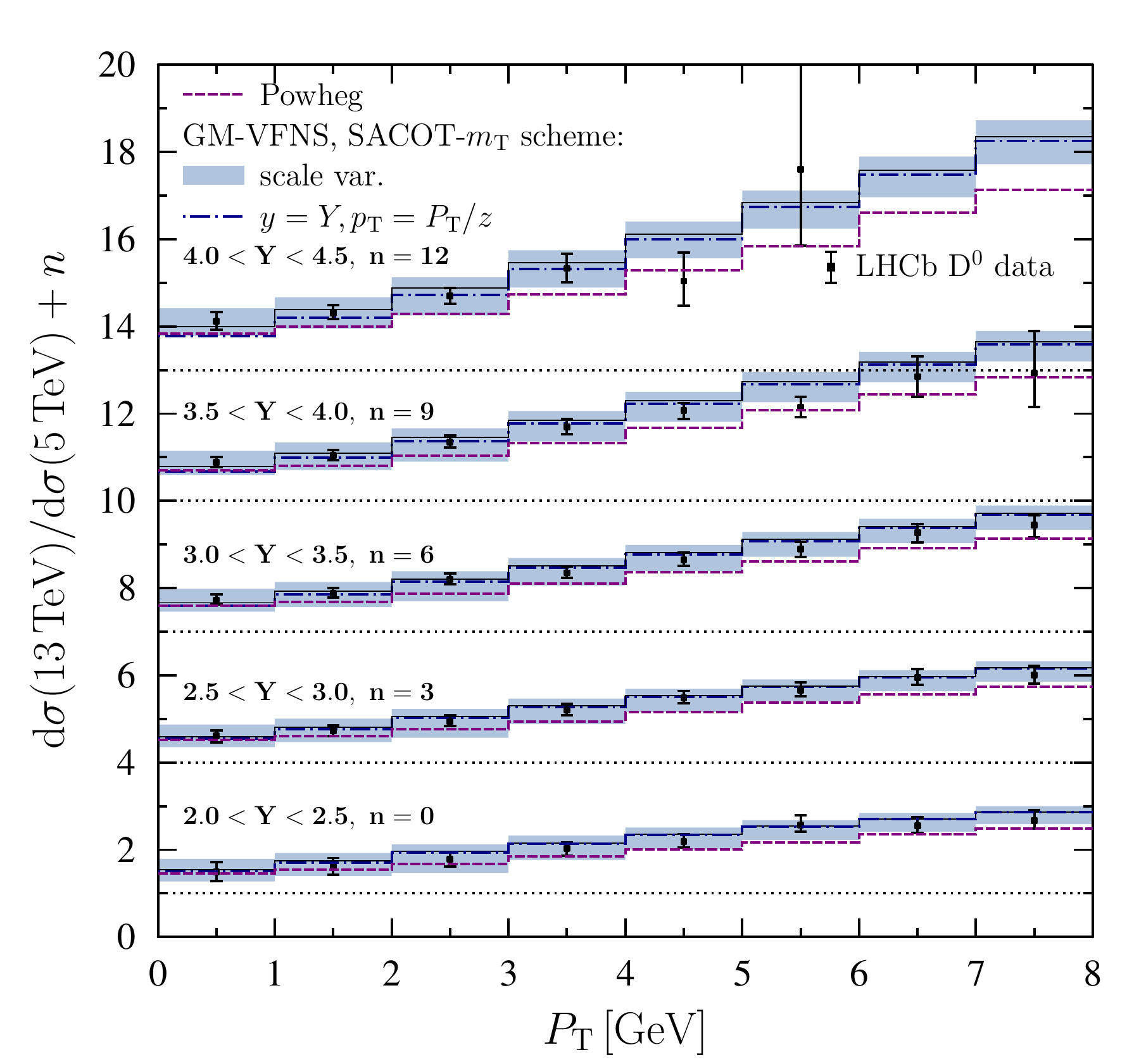}
\caption{
As the upper panel of Figure~\ref{fig:LHCbratios}, but without the PDF uncertainties. Also the predictions ignoring the mass dependence in the fragmentation variable $z$ are shown for comparison (blue dashed-dotted lines). 
} 
\label{fig:LHCbratios2}
\end{figure}

Figure~\ref{fig:LHCbratios} presents ratios of cross sections measured by LHCb at different $\sqrt{s}$. Here, a significant part of the theory uncertainties cancel and, sure enough, the central predictions of the both considered methods describe the data rather well even at low $P_{\rm T}$. However, the \textsc{Powheg} results are systematically below the GM-VFNS predictions in most of the cases, best visible in the $(\sqrt{s}=13 \, {\rm TeV})/(\sqrt{s}=5 \, {\rm TeV})$ panel. That is, the $\sqrt{s}$ dependence is stronger in the GM-VFNS calculation. We believe that this hierarchy follows mainly from the presence (absence) of gluon fragmentation in GM-VFNS (\textsc{Powheg}+\textsc{Pyhtia}) approach, similarly as in the case of absolute cross sections. As the \textsc{Powheg} generates events where the $Q\overline{Q}$ pair is produced in the hard process or from the hardest emission, part of the increased phase-space for parton shower due to increased $\sqrt{s}$ is not in use for heavy-quark production. This is consistent also with the observation that the difference between the GM-VFNS and \textsc{Powheg}+\textsc{Pythia} cross sections decreases, though slowly, towards lower $\sqrt{s}$. In Figure~\ref{fig:LHCbratios2} we show the $(\sqrt{s}=13 \, {\rm TeV})/(\sqrt{s}=5 \, {\rm TeV})$ case without PDF errors and including also the predictions obtained with the zero-mass version of the fragmentation variable ($z=P_{\rm T}/p_{\rm T}$, $y=Y$) to estimate the theoretical bias of this origin. We observe that the differences between our default definition and the zero-mass version are still clearly smaller than the scale uncertainties.

\begin{figure}[htb!]
\centering
\includegraphics[width=0.66\textwidth]{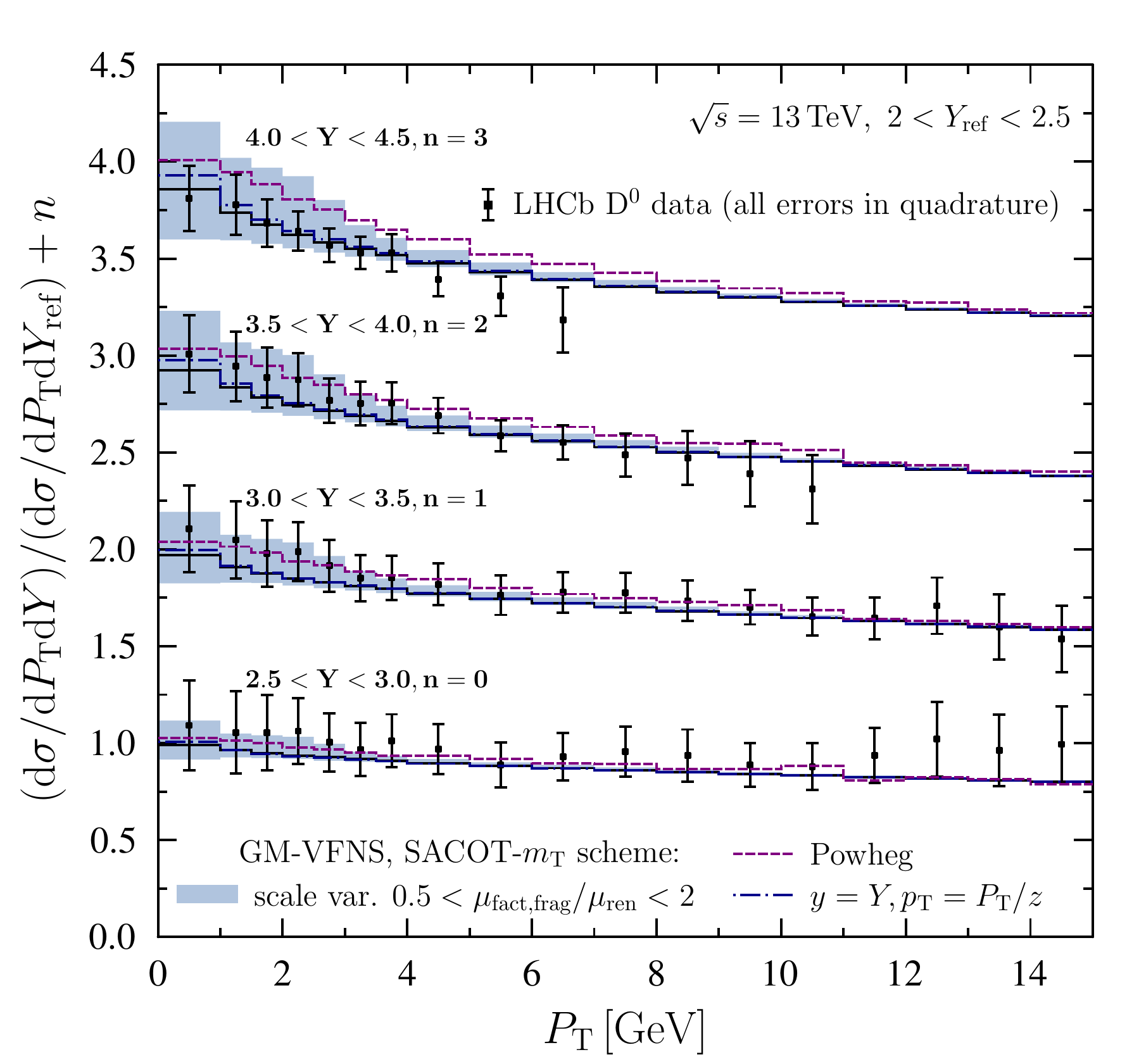}
\caption{
Cross sections at $\sqrt{s}=13\,{\rm TeV}$ normalized to the rapidity bin $2.0<Y_{\rm ref}<2.5$. Black lines are the central GM-VFNS results, and the light blue band corresponds to the scale variation. The dark blue dashed-dotted lines are the predictions with zero-mass fragmentation variable, and purple dashed ones indicate the \textsc{Powheg}+\textsc{Pythia} prediction. The data are from Ref.~\cite{Aaij:2015bpa}.
} 
\label{fig:LHCbratios3}
\end{figure}

Another observable that has been discussed in the recent literature \cite{Zenaiev:2015rfa,Gauld:2016kpd} is the normalized cross section 
\begin{equation}
\frac{{\rm d}\sigma/({\rm d}P_{\rm T}{\rm d}Y)}{{\rm d}\sigma/({\rm d}P_{\rm T}{\rm d}Y_{\rm ref})} \,,
\end{equation}
where $Y_{\rm ref}$ is a fixed reference rapidity. Also here, large part of the theory uncertainties cancel upon taking the ratio. Our calculations for this quantity within the LHCb acceptance at $\sqrt{s}=13\,{\rm TeV}$ in GM-VFNS framework are presented in Figure~\ref{fig:LHCbratios3} taking $2<Y_{\rm ref}<2.5$. We also compare with the \textsc{Powheg}+\textsc{Pythia} approach and with the GM-VFNS predictions using the zero-mass fragmentation variable. The shown LHCb data points have been formed from the absolute cross sections adding all the uncertainties in quadrature (as if they were uncorrelated). Again, the scale variation appears more significant than the effect induced by using the zero-mass fragmentation variable ($z=P_{\rm T}/p_{\rm T}$, $y=Y$). The \textsc{Powheg}+\textsc{Pythia} results are systematically above the GM-VFNS ones. As earlier, this seems to follow from the absence of gluon fragmentation in \textsc{Powheg}+\textsc{Pythia}: The ``missing'' contributions from gluon fragmentation are relatively larger for $2<Y_{\rm ref}<2.5$ than for the more forward bins (since the phase space is larger for $2<Y_{\rm ref}<2.5$), and thus the ratio is higher than in GM-VFNS.

\section{Summary}
\label{sec:Summary}

In the present article, we have introduced a novel implementation of the GM-VFNS for hadroproduction of heavy-flavoured mesons. Here, the novelty amounts to a specific definition of scheme, SACOT-$m_{\rm T}$, which retains the kinematics of heavy quark-antiquark pair production also in contributions where the heavy-flavoured meson formally comes from light-flavour fragmentation or initial-state heavy quarks. As we have explained, this is physically a natural choice as the origin of these contributions is in the heavy quark-antiquark pair production, and as such it is analogous to the SACOT-$\chi$ scheme in deeply inelastic scattering. Within the SACOT-$m_{\rm T}$ scheme, it is possible to compute the heavy-flavoured meson spectra down to $P_{\rm T}=0$ with arbitrary choices for renormalization, fragmentation, and renormalization scales. In earlier works presented in the literature, a finite $P_{\rm T} \rightarrow 0$ limit could only be achieved by setting the scales in a particular way. Comparisons with the available D$^0$-meson data from LHCb and ALICE collaborations indicate that our calculation in SACOT-$m_{\rm T}$ scheme performs well, though it must be admitted that the theoretical uncertainties at low $P_{\rm T}$ are significant. Here, it is not only the scale, PDF, and FF uncertainties that matter, but also the scheme dependence and ambiguities in defining the fragmentation variable $z$ in the presence of finite heavy-quark and heavy-flavoured meson masses. In particular, as we have shown, the latter can have a significant impact on the shape of the absolute spectra at low $P_{\rm T}$. Within our definition of the fragmentation variable we have found, however, that the mass effects are suppressed in the forward direction,  especially within the LHCb acceptance. Nevertheless, a particular choice of fragmentation variable may still bias the usage of low-$P_{\rm T}$ D-meson production e.g. as a constraint for PDFs. We note that this type of uncertainty is not inherent only to the GM-VFNS approach but the very same ambiguity is there also in FFNS calculations when scale-independent FFs are used for the $c\rightarrow D$ transition. We have also shown that in all considered rapidities there is a sizeable contribution from large-$x_2$ region which, as we have argued, appears to originate from gluon fragmentation. Thus, estimates based on fixed-order ${\rm c\overline{c}}$ pair production overstate the small-$x$ sensitivity of inclusive D-meson production. Though formally higher order in strong coupling, the effect of gluon fragmentation is numerically large and seems to explain why FFNS-based calculations are typically a factor of two below the LHC data. In addition, we have observed that an approach (\textsc{Powheg}+\textsc{Pythia}) neglecting large part of the gluon fragmentation deviates systematically from the GM-VFNS predictions also in the case of normalized cross sections and ratios across different $\sqrt{s}$. The size of these systematic differences competes or exceeds the scale uncertainty of GM-VFNS, though the scale uncertainties in the \textsc{Powheg}-based setup are presumably somewhat larger also for these observables. In the future, we plan to extend the SACOT-$m_{\rm T}$ scheme also to the case of intrinsic/fitted charm, B-meson production, as well as nuclear collisions.

\section*{Acknowledgments}

This research was supported by the Academy of Finland, Projects 297058 and 308301, as well as by the Carl Zeiss Foundation. The Finnish IT Center for Science (CSC) is acknowledged for super-computing time under the Project jyy2580. In addition, the authors acknowledge support by the state of Baden-Württemberg through bwHPC, and thank Kari J.~Eskola for discussions in the early stages of this work. The authors also wish to thank the JHEP Referee for comments and suggestions which have helped us to improve the paper.

\end{document}